\documentclass[twocolumn]{aastex631}

\usepackage{xspace}
\usepackage{hyperref}
\usepackage{amsmath}

\newcommand{\HST}{{\it HST}\xspace}
\newcommand{\RomanST}{{\it Roman}\xspace}
\newcommand{\RomanSpelled}{{\it Nancy Grace Roman Space Telescope}\xspace}
\newcommand{\HubbleSpaceTelescope}{{\it Hubble Space Telescope}\xspace}

\newcommand{\scatter}{dispersion\xspace}
\newcommand{\Scatter}{Dispersion\xspace}

\newcommand{\committee}{CCS Committee\xspace}
\newcommand{\committeesdef}{Core Community Survey (CCS) Committees\xspace}
\newcommand{\eminus}{e$^-$\xspace}
\newcommand{\numsurveys}{1,000\xspace}
\newcommand{\interlaced}{interlaced\xspace}
\newcommand{\slewsettle}{50s\xspace}
\newcommand{\CCSsurvey}{baseline CCS survey\xspace}
\newcommand{\ReferenceSurvey}{Reference survey\xspace}

\newcommand{\ztarget}{z_{\mathrm{target}}}

\newcommand{\ensuretext}[1]{\ifmmode\text{#1}\else#1\fi}

\newcommand{\SNRSUM}{SNR~Sum\xspace} 
\newcommand{\ang}{\ensuretext{\AA}\xspace}

\begin{document}
\title{Fishing for the Optimal \RomanST High Latitude Time Domain Survey:\\ Cosmological Constraints for 1,000 Possible Surveys}

\newcommand{\uhawaii}{\affiliation{Department of Physics and Astronomy, University of Hawai`i at M{\=a}noa, Honolulu, Hawai`i 96822}}
\newcommand{\stsci}{\affiliation{Space Telescope Science Institute, 3700 San Martin Drive Baltimore, MD 21218, USA}}
\newcommand{\lbnl}{\affiliation{E.O. Lawrence Berkeley National Laboratory, 1 Cyclotron Rd., Berkeley, CA, 94720, USA}}
\newcommand{\duke}{\affiliation{Department of Physics, Duke University, Durham, NC 27708, USA}}
\newcommand{\UMBC}{\affiliation{University of Maryland, Baltimore County, Baltimore, MD 21250, USA}}
\newcommand{\Goddard}{\affiliation{NASA Goddard Space Flight Center, Greenbelt, MD 20771, USA}}
\newcommand{\UPenn}{\affiliation{Department of Physics and Astronomy, University of Pennsylvania, Philadelphia, PA 19104, USA}}
\newcommand{\KICP}{\affiliation{Kavli Institute for Cosmological Physics, University of Chicago, Chicago, IL 60637, USA}}
\newcommand{\UCAA}{\affiliation{Department of Astronomy and Astrophysics, University of Chicago, Chicago, IL 60637, USA}}
\newcommand{\berkeley}{\affiliation{Department of Physics, University of California Berkeley, Berkeley, CA 94720, USA}}

\newcommand{\FRR}{\ensuremath{F062}\xspace}
\newcommand{\FZ}{\ensuremath{F087}\xspace}
\newcommand{\FY}{\ensuremath{F106}\xspace}
\newcommand{\FJ}{\ensuremath{F129}\xspace}
\newcommand{\FH}{\ensuremath{F158}\xspace}
\newcommand{\FF}{\ensuremath{F184}\xspace}
\newcommand{\FK}{\ensuremath{F213}\xspace}

\author[0000-0001-5402-4647]{David~Rubin}
\uhawaii
\lbnl

\author{Greg Aldering}
\lbnl

\author{Andy Fruchter}
\stsci

\author{Llu\'is Galbany}
\affiliation{Institute of Space Sciences (ICE, CSIC), Campus UAB, Carrer de Can Magrans, s/n, E-08193 Barcelona, Spain}
\affiliation{Institut d'Estudis Espacials de Catalunya (IEEC), E-08034 Barcelona, Spain}

\author[0000-0002-0476-4206]{Rebekah~Hounsell}
\UMBC
\Goddard

\author{Rick Kessler}
\KICP
\UCAA

\author{Saul Perlmutter}

\lbnl
\berkeley

\author{Ben Rose}
\affiliation{Department of Physics and Astronomy, Baylor University, Waco, TX 76706, USA}

\author[0000-0003-2764-7093]{Masao~Sako}
\UPenn

\author[0000-0002-4934-5849]{Dan~Scolnic}
\duke

\author{Jannik Truong}
\berkeley

\author{the Roman Supernova Cosmology Project Infrastructure Team}

\begin{abstract}

The upcoming \RomanSpelled is set to conduct a generation-defining SN~Ia cosmology measurement with its High Latitude Time Domain Survey (HLTDS). However, between optical elements, exposure times, cadences, and survey areas, there are many survey parameters to consider. This work was part of a \RomanST Project Infrastructure Team effort to help the Core Community Survey (CCS) Committee finalize the HLTDS recommendation to the \RomanST Observations Time Allocation Committee. We simulate 1,000 surveys, with and without a conservative (volume-limited) version of the Vera C. Rubin Observatory Deep Drilling Field SNe~Ia, and compute Fisher-matrix-analysis Dark Energy Task Force Figures of Merit (FoM, based on $w_0$-$w_a$ constraints) for each. We investigate which survey parameters correlate with FoM, as well as the dependence of the FoM values on calibration uncertainties and the SN \scatter model. The exact optimum depends on the assumed \scatter model and whether Rubin DDF SNe~Ia are also considered, but $\sim 20\%$ time in prism, $\sim$~30--40\% time in Wide imaging and the remainder in Deep imaging seems most promising. We also advocate for ``\interlaced'' cadences where not every filter is used in every cadence step to reduce overheads while maintaining a good cadence and increasing the number of filters compared to the \citet{Rose2021} reference survey (the prism has proportionately lower overheads and can be used for each cadence step). We show simulated light curves and spectra for the baseline HLTDS CCS recommendation and release distance-modulus covariance matrices for all surveys to the community.

\end{abstract}

\keywords{Type Ia supernovae, Cosmology, Dark energy, Surveys}

\section{Introduction and General Approach} \label{sec:intro}

The accelerated expansion of the universe was discovered more than 25 years ago \citep{Riess1998, Perlmutter1999}, however the cause remains mysterious. Although much progress in narrowing down the phenomenology of the acceleration has been made with ground-based facilities and the \HubbleSpaceTelescope (e.g., \citealt{Riess2004, Astier2006, Kessler2009, Suzuki2012, Betoule2014, Scolnic2018, Alam2021, Brout2022, Rubin2023UNITY, DESCollaboration2024, DESICollaboration2024}), 
the 2010 Decadal Survey recommended a space-based mission with dedicated surveys for dark energy and dark matter, and exoplanets \citep{Decadal2010}.

The \RomanSpelled (\RomanST) came out of that recommendation and is the next NASA flagship mission, with launch as soon as October 2026. Its five-year prime mission will be a mixture of collaboratively defined surveys suggested by the \committeesdef and General Observer (GO) programs. One survey that recently underwent definition is the High Latitude Time Domain Survey (HLTDS, \citealt{ObservationsTimeAllocationCommittee2025}). With \RomanST's large 0.281 deg$^2$ optical--NIR camera, low observing overheads, and both imaging and spectroscopic capabilities, the HLTDS can deliver a generation-defining SN cosmology measurement using $\mathcal{O}$(10,000) SNe~Ia. The work described in this paper was part of a Roman Project Infrastructure Team effort to help optimize the HLTDS for SN cosmology while respecting constraints that come from balancing against other science goals.

This optimization is not straightforward, as the number of \RomanST HLTDS possibilities is very large (discussed more in Section~\ref{sec:Surveys}). There are seven optical elements to consider: \FRR, \FZ, \FY, \FJ, \FH, \FF, and the prism.\footnote{The \FK and grism may also be used, but probably not cadenced over large areas, so are not considered here.} Generally, surveys have different tiers that trade area against depth \citep{York2000, CFHTLS}, so each optical element could be part of two to three tiers with different exposure times per tier. In addition, ground-based optical observations, e.g., with the Vera C. Rubin Observatory \citep{LSSTScienceCollaboration2009, Ivezic2019} could be considered. Each tier covers a certain amount of area with a certain cadence (and the cadences can also be different between optical elements).\footnote{In addition, the assumptions about the instrument and supernovae could be varied, e.g., SN rates, number of external nearby SNe assumed, or calibration uncertainties. The survey duration could be varied; the survey nominally takes 0.5~years, but could take longer or shorter.}

Our primary approach is thus to randomly sample areas of survey parameter space that are considered promising (although we also did additional targeted exploration over small numbers of dimensions at the request of the \committee). This work analyzed \numsurveys simulated surveys under different assumptions (possibly the largest such set of proposed surveys in one work). We compute relative Figure of Merit (FoM) values for each survey (discussed more in Section~\ref{sec:Fisher}) related to the cosmological constraining power of each. This is a key point to keep in mind for this paper: these are intended to be {\it relative} FoM values to rank order the surveys. As noted above, the true FoM value will depend on some parameters (like SN rates, calibration uncertainties, or the supernova \scatter model) that we will not know with precision until the data start coming in. But as we will see, one can still select surveys that perform better than others under a range of assumptions.

There is no one-size-fits-all approach to survey optimization. Frequently, catalog-level simulations are produced for a variety of survey strategies and passed through a simulated analysis \citep{Bernstein2012, Hounsell2018}. Another common approach is to construct Fisher matrices for the survey (for a good introduction to Fisher matrices, see \citealt{Coe2009}). If these matrices fit the simulated fluxes for each object, then they can approximate a true optimal simultaneous analysis of the data, accounting for supernova magnitude and extinction, calibration uncertainties, the training of the mean SN model, and cosmological parameters (c.f., \citealt{Astier2011, Astier2014}; other notable Fisher matrix analyses are \citealt{TheLSSTDarkEnergyScienceCollaboration2018, Gris2023}).\footnote{A true simultaneous analysis also has the maximum potential for self calibration \citep{Kim2006}.}\textsuperscript{,}\footnote{Some Fisher-matrix approaches only approximate the results of the light-curve fitting rather than fitting fluxes and/or do not treat some of these terms \citep{Miknaitis2007, Rubin2020}. We thus do not consider that approach here as it is less accurate.} In principle, Fisher-matrix analyses can treat selection effects and non-Ia contamination at lowest order, but we do not believe these will be dominant in surveys of this quality \citep{Abdelhadi2024}, so we opt to simplify and exclude these. Fisher-matrix optimization is generally less computationally intensive (for example, when fitting light-curves, the light-curve fitting is only done at first order rather than full nonlinear light-curve fit in a simulated analysis). For the above reasons, we opted to use such a Fisher-matrix approach for our \numsurveys simulated surveys.

This paper is organized as follows. Section~\ref{sec:Surveys} describes the parameters that were varied to make each survey. Section~\ref{sec:Simulations} describes the survey simulations. Section~\ref{sec:Fisher} describes the Fisher-matrix calculations, including the systematic uncertainties included. Section~\ref{sec:Results} describes the results. Finally, Section~\ref{sec:Conclusion} concludes with a summary of the main points.

\section{Surveys Considered} \label{sec:Surveys}

\begin{figure}[htbp]
\begin{center}
\includegraphics[width = 0.4\textwidth] {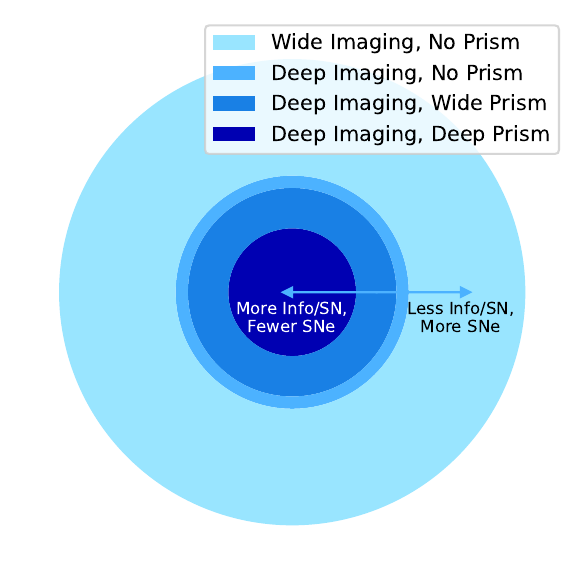}
\caption{An illustration of a possible arrangement of ``wedding-cake'' tiers in the simulated surveys. The prism tiers are independent of the imaging tiers. In general, deeper prism/imaging data are taken together so that the SNe towards the center have the most complete data (deepest imaging, deepest prism). In general, one wants deeper tiers inside wider tiers so that a SN that falls out of the Deep tiers due to edge effects (\RomanST's focal plane cannot perfectly fill a circle, and the whole tiling pattern rotates $\sim 1^{\circ}$ per day to keep the solar panels facing the Sun) lands in the Wide tiers instead of being lost completely.}
\label{fig:tiers}
\end{center}
\end{figure}

Each survey is parameterized in terms of the parameters for two imaging tiers:\footnote{We did not find much benefit to three-imaging-tier surveys for \RomanST, so none of those surveys are presented.} ``Wide'' and ``Deep,'' and three prism tiers: ``Wide,'' ``Deep,'' and none (just imaging). The tiers are assumed concentric to minimize edge effects, i.e., Deep footprint inside Wide footprint, so that if a SN falls out of the Deep as \RomanST rotates it is at least covered by the Wide \citep{Rose2021}. As much of the prism as possible is inside the Deep-imaging tiers, i.e., even the Wide-tier prism is coupled to the deepest possible imaging to extract the most information about the SNe observed in the prism. (The Wide-imaging tier is also better able to use ground-based spectroscopy is more feasible due to its lower redshift range.) This is illustrated in Figure~\ref{fig:tiers}.

Each tier has:
\begin{itemize}
    \item A fraction of time for the tier out of the total survey time (nominally, six months total time).
    
    \item A cadence in observer-frame days which can depend on the optical element (filters or prism), described in Section~\ref{sec:cadence}.

    \item A choice of filters from \FRR, \FZ, \FY, \FJ, \FH, \FF, and the prism, described in Section~\ref{sec:filtchoice}.

    \item A targeted redshift ($\ztarget$) to set the imaging exposure times, described in Section~\ref{sec:ztarget}.

\end{itemize}

\subsection{Cadence} \label{sec:cadence}

 Iterating with the \committee, it became clear that \interlaced cadences yielded a good balance of sampling and area. Here, one does half the selected imaging filters in every other cadence step. The exposure times have to increase (discussed in Section~\ref{sec:ztarget} around Equation~\ref{eq:exptime}) to maintain the same average ``depth per day,'' but the \interlaced cadence still covers more area than an ``all-filters-every-visit'' cadence because these longer exposures have proportionately lower read noise and slew overheads \citep{Rubin2023Cadence}.
 
 The \committee also asked to have one filter always observed (i.e., observed at every cadence step and thus not \interlaced), ideally one of the bluer filters as transients evolve faster in the blue. For example, \citet{Rose2021} suggested \FRR, \FZ, \FY, \FJ for the Wide-imaging tier with a cadence of five days. So instead, we might do $\FRR + \FY$, then five days later, $\FRR + \FZ + \FJ$. (All filter choices considered here are described in Section~\ref{sec:filtchoice}.)

 In this work, we considered same-filter cadences of $\{4, 6, 8, 10\}$ days for the Wide-imaging tier and $\{6, 8, 10, 12\}$ for the Deep-imaging tier; in other words, half the imaging filters every $\{2, 3, 4, 5\}$ days for Wide and $\{3, 4, 5, 6\}$ days for Deep. As our Fisher-matrix analysis does not simulate classification (or SN~Ia subclassification), setting the upper limit for the cadence must be done by hand and verified with more detailed studies \citep{Abdelhadi2024, Rose2025}.

In general, the prism can tolerate a much faster cadence than the imaging without loss of efficiency as the prism exposures are much longer so the prism is photon-noise-dominated (not read-noise-dominated as some of the short imaging exposures are, \citealt{Rubin2023Cadence}). The prism depths are 900~s for the Wide and 3600~s for the Deep and assume a 5-day cadence (c.f., \citealt{Rose2021} and see the optimization in \citealt{Rubin2022}). A short cadence of a few rest-frame days helps with training the SN~Ia model. In practice one would sync up the prism and the imaging observations to save on slew time (e.g., for a 10-day \interlaced cadence, do the prism every 5 days). Also, prism exposures should be broken into shorter dithered exposures of perhaps 450 seconds to sample sub-pixel information and span the chip gaps. However, we do not simulate this here as the exposure times are much longer than the time it takes to dither and the increased read noise from dithering is not very relevant given the high prism sky background.

\subsection{Filter Choice} \label{sec:filtchoice}

\newcommand{\HowManyFilters}{We consider four possibilities for the Wide-imaging tier and five possibilities for the Deep-imaging tier.\xspace}

 \begin{figure*}[htbp]
\begin{center}
\includegraphics[width = \textwidth] {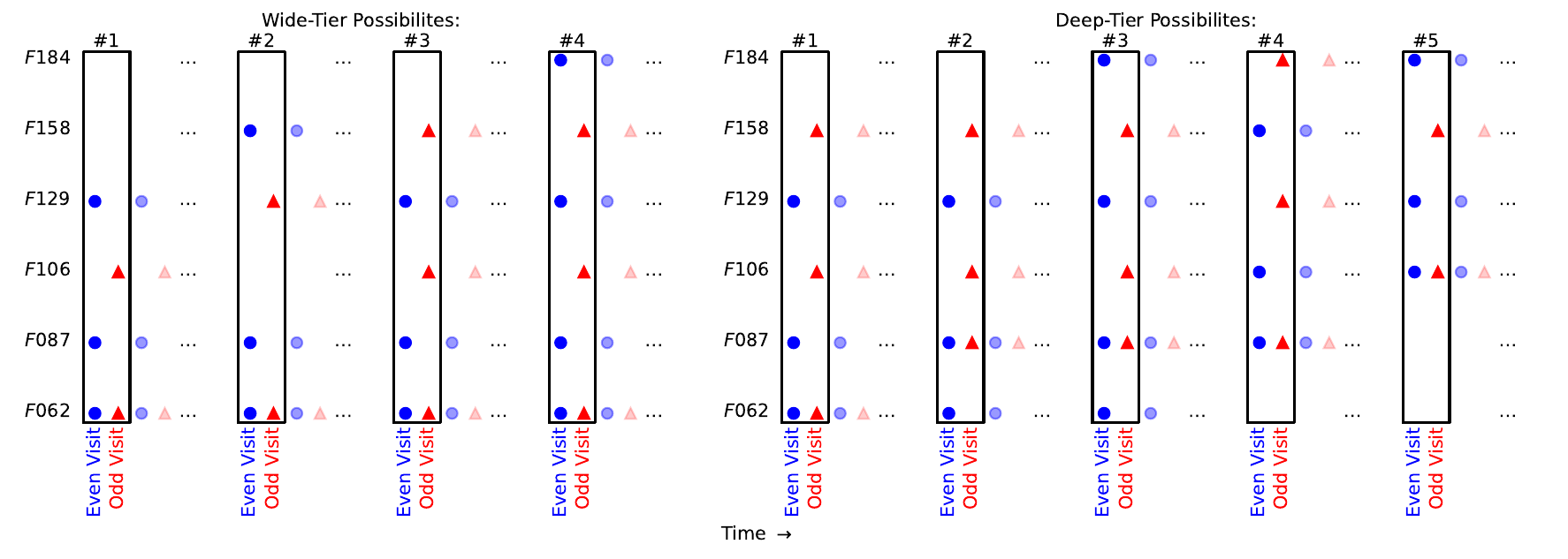}
\caption{Filter possibilities considered in this work. \HowManyFilters Most filters are only taken every other visit, but one bluer filter is always observed. The repeated pattern is boxed for each possible combination of filters.}
\label{fig:filters}
\end{center}
\end{figure*}

The \cite{Rose2021} proposed surveys had \FRR, \FZ, \FY, and \FJ in the Wide-imaging tier, and \FY, \FJ, \FH, \FF, and the prism in the Deep tier, but we will see that the \committee recommendation of adding filters is superior, and becomes feasible with the \interlaced cadences. We considered the filter combinations shown in Figure~\ref{fig:filters}; the single always-observed filters discussed above are visible. \HowManyFilters The Wide-imaging-tier strategies always include the bluest filters (\FRR and \FZ), which generally have shorter exposure times (Table~\ref{tab:exposuretime}) and the possibilities span which of the redder filters to include. The Deep-imaging-tier strategies are generally aimed at higher-redshift SNe and consider dropping either the bluest filters (\FRR or \FZ) or the expensive \FF.

\subsection{Exposure Times} \label{sec:ztarget}

Our exposure times for targeting a certain redshift $\ztarget$ are given in Table~\ref{tab:exposuretime} and are computed using the following constraints.
    \begin{itemize}
        \item The surveys target an imaging S/N per point at maximum light of
\begin{equation} \label{eq:exptime}
    {\rm S/N} = 10 \sqrt{\frac{\mathrm{cadence}}{(2.5\ \mathrm{days})(1 + \ztarget)}} \;
\end{equation}
(c.f., \citealt{Miknaitis2007} Equation~4) for the median SN/host-galaxy background at the targeted redshift.
So in a $\pm 5$ rest-frame day range around maximum, there will be an average S/N of 20 for the median SN at that redshift. 
The effective timescale of a SN~Ia is about 20 days in the rest-frame (depending on wavelength),\footnote{We evaluate this timescale by computing $\sum_{\mathrm{phase}\ i} [\mathrm{flux}_i/\mathrm{maximum}]^2$ using the \citet{Hsiao2007} light-curve templates, c.f., our Equation~\ref{eq:SNRSum}.} so over the full light curve, a total S/N of $\approx 10\sqrt{8} = 28$ will be reached in each filter (again for the median SN), a high quality measurement that is comparable to the per-band \scatter for SNe Ia \citep{Pierel2022}. For comparison, \citet{Astier2011} recommended 0.025~magnitudes measurement uncertainty in amplitude per band for rest-frame $BVR$ or S/N $\approx 43$ and \citet{Gris2023} required color uncertainty $< 0.04$~mag.
    \item Imaging exposure times must not decrease with increasing redshift. This requirement is due to some filters coming in and out of spectral features with redshift and thus increasing and decreasing exposure times. Another way to state this is that targeting, e.g., $z=1.7$ also targets $z=0.1,~\ldots,~1.7$, and the exposure times are set to the maximum for the redshift range.\footnote{For readers used to CCDs: a useful feature of the HAWAII CMOS that \RomanST uses is that each pixel can be continuously nondestructively read during the exposure, so that one does not have to worry about saturating on brighter SNe during long exposures unless those SNe are brighter than $\sim 15$~AB. It is unlikely that even one SN this bright will be observed.}
    \item Imaging exposure times stop increasing with redshift when filter effective wavelength goes bluer than 3,500\ang in the rest-frame. SNe~Ia are generally faint in the UV and it is unclear if deep rest-frame UV observations help with standardization. So this requirement keeps the bluest filters from dramatically increasing their exposure times with no clear benefit when exposing for high redshift.
    \item We do not consider any exposure times shorter than 60s, as even at 60s the overheads and read noise significantly degrade the S/N compared to longer exposures. This is reduced from 100s minimum in \citet{Rose2021}, partially because the 70s slew+settle overheads assumed there have since been reduced to $\sim \mathrm{\slewsettle}$.
    \end{itemize}

\section{Simulations} \label{sec:Simulations}

The simulation framework used in this work is based on the framework developed for \citet{Rubin2020, Rubin2022} and is available at \url{https://github.com/rubind/wfirst-sim}. In short, it uses SN volumetric rates based on \citet{Rodney2014} to draw a given number of SNe, their redshifts (quantized in bins of 0.05), and their times of maximum inside the survey window of two years. It draws the SALT2/3 \citep{Guy2007, Kenworthy2021, Pierel2022} light-curve parameters $m_B$, $x_1$, $c$ for each SN using population fits to the lower-redshift (more complete) portion of the SDSS and SNLS SNe from \citet{Betoule2014}. The host-galaxy background is based on SED fits to the surface brightnesses inside a $0\farcs1$-radius aperture at the location of the \HST GOODS/MCT SNe \citep{Riess2004, Riess2007, Riess2018} and so should be appropriate for \RomanST which is also a 2.4~m space telescope observing $z \sim 1$ host galaxies. Each surface brightness (in $F435W$ through $F160W$) used to perform an SED fit. We convert to log flux and compute the mean (and its dispersion) and the first principal component (and its dispersion). We randomly sample from this mean and dispersion (assumed Gaussian) from these components for the simulations. Thus, each SN has a semi-realistic host galaxy background as a function of wavelength.\footnote{For redshifts much lower than 1, this prescription is probably incorrect in detail. However, the S/N will be so high that it does not matter in practice.}

We find the statistical uncertainty on FoM is $\sim2\%$ realization-to-realization, unsurprising for a survey that measures $\mathcal{O}$(10,000)~SNe~Ia and thus has $>1\%$ statistical fluctuations in SN number counts as a function of redshift. We suppress most of this uncertainty for this work by simulating the number counts of SNe as a function of redshift exactly, without Poisson noise. Some FoM statistical uncertainty remains, as $m_B/x_1/c$ fluctuations or whether a SN explodes near the survey start/end can bring a SN in or out of the sample. Of course, the real survey will have the full statistical uncertainty.

We use \texttt{SNCosmo} \citep{Barbary2016a} to produce true SN SEDs for each date of the survey for each active SN using SALT3 NIR \citep{Pierel2022}.

For the imaging, we simulate a small pixelized image using a \texttt{WebbPSF} PSF with a random subpixel centroid for each filter/epoch and convolved with an inter-pixel capacitance kernel of 0.02 for the four nearest neighbors (and thus 0.92 for the central pixel). We add appropriate sky background \citep{Aldering2002}, an assumed 0.015~e$^-$/pix/s scattered light, thermal background (the \citealt{Rubin2020} model, assuming a 264K primary and secondary), and host-galaxy background, all assumed spatially constant over the PSF size with values given in Table~\ref{tab:perfiltvals}. Then we add read noise \citep{Rauscher2007} of 16~\eminus per read, a 5~\eminus floor, and Poisson noise. We add all noise sources in quadrature, although this is not quite correct in detail:
\begin{equation}
\mathrm{Noise} = \sqrt{12\cdot 16^2 \frac{N - 1}{N(N+1)} + 5^2 + \mathrm{Counts}} \;,
\end{equation}
where the number of reads $N$ is the exposure time divided by the time per full readout (``frame time'') of 3.04 seconds.
We perform a simple optimal (PSF-weighted) extraction to obtain flux and uncertainty. We assume perfect subtraction of the host-galaxy light (other than the Poisson-noise contribution) as forward modeling performs well on simulated data \citep{Rubin2021}. We simulate chip gaps by assuming a 91\% fill factor\footnote{The pointing offsets we assume are 0.82034$^{\circ}$ in the long focal-plane direction by 0.37628$^{\circ}$ in the short focal-plane direction for a total of 0.3087 square degrees per pointing with 0.281 square degrees filled by detectors. Any observation landing in a chip gap is thrown out. Finally, we assume the chip gaps are randomly distributed across SNe, time, and filters, which is mostly correct since the whole survey area will rotate $\sim 1^{\circ}$/day to keep Roman's solar panels pointed at the Sun. In practice, one should perform a chip-gap-spanning dither between filters so that multiple filters cannot be missing in the same epoch.} and assume 1\% bad pixels (randomly distributed) in the small image patches.

\begin{deluxetable}{lcc}
\label{tab:perfiltvals}
\tablecaption{Per-element simulation values. Background includes zodiacal, thermal, dark, and scattered light. The prism background is zodiacal-dominated and is thus more similar SN-to-SN than are the imaging backgrounds, so the assumed host-galaxy background is included in this table only for the prism (the imaging host-galaxy backgrounds are discussed in the text).}
    \tablehead{\colhead{Element}	&	\colhead{Background (e$^-$/pix/s)} &	\colhead{AB ZP}}
\startdata
\FRR	&	0.330	&	26.6187	\\
\FZ	&	0.331	&	26.3027	\\
\FY	&	0.372	&	26.3552	\\
\FJ	&	0.374	&	26.3538	\\
\FH	&	0.362	&	26.3765	\\
\FF	&	0.371	&	25.9129	\\
\FK	&	4.804	&	25.864	\\
Prism	&	1.200	&	\nodata	
\enddata
\end{deluxetable}

The prism data are handled similarly, except that the PSF is highly chromatic (both because different wavelengths are dispersed by different amounts and because the aberrations are worse). We take the prism dispersion from the Project (\texttt{GRISM\_PRISM\_Dispersion\_190510.xlsx}) and convert it to a mean wavelength for each pixel along the trace.\footnote{We also exclude the prism PSF along the dispersion outside a five-pixel range centered on the brightest pixel. This slightly reduces the S/N, but in a real analysis (not the Fisher analysis here), the eliminated signal would introduce correlations that would have to be modeled out (e.g., \citealt{Bolton2010}). So probably the additional S/N is an illusion. In any case, clipping the PSF and lowering the S/N is conservative.} For the prism, the noise from the host-galaxy background is essentially irrelevant (backgrounds are dominated by zodiacal light) so we simply increase the zodiacal background by 5\% (assuming that the host is comparable to zodiacal background over about 10 pixels and the prism trace is 200 pixels long, \citealt{Rubin2022}). As in the imaging host-galaxy noise model, this model will be wrong at low redshift, but again the S/N will be very high in this case so this is unimportant. We assume perfect subtraction of the host-galaxy light (other than the Poisson-noise contribution), just as we do in the imaging.\footnote{As each prism roll angle will be visited roughly twice over the two-year survey, removing host-galaxy light with a 2D subtraction of the two visits would incur a $\sqrt{2}$ scale increase on the noise. Avoiding this increase will require fitting a 3D ``scene model'' constructed using data from a range of roll angles to break spatial/spectral degeneracy; \citet{Astraatmadja2025} discuss this in detail and demonstrate its feasibility so we do not take the $\sqrt{2}$ factor.} We generate the prism data at native resolution but bin it to 25~bins (in log wavelength) for the Fisher calculation (Section~\ref{sec:Fisher}) to reduce RAM requirements.\footnote{As described in Section~\ref{sec:Fisher}, the mean SN model is only computed at moderate spectral resolution. In contrast, the native prism spectra have $\sim 200$ wavelengths over the range 7500\ang to 18000\ang.}

We also simulate Vera C. Rubin Observatory observations of lower-redshift SNe. We always assume 800 $z < 0.1$ nearby SNe. We also run each survey twice, with and without $\sim $3,500 SNe from the deep drilling fields (DDF). We take only a volume-limited subset of Rubin SNe ($z<0.5$), so we do not have to worry about Rubin selection effects. To be conservative, we assume a single visit in $griz$ every four nights ($5\sigma$ depths of $g=24.47$, $r=24.16$, $i=23.4$, and $z=22.23$). Also to be conservative, we do not assume the Rubin observations overlap the \RomanST HLTDS. However, we do assume the full 10 years, with an average of four extragalactic DDF fields observed for 0.4 years every year over the 10-year survey (16 field-years total).

\section{Fisher-Matrix Cosmology Analysis} \label{sec:Fisher}

\newcommand{\designmatrix}{\ensuremath{\left( \tfrac{\partial \mathrm{Model}_u}{\partial \mathrm{Param}_v} \right)}\xspace}

The central idea of a Fisher-matrix calculation is to approximate a full analysis by linearizing the model and assuming all uncertainties are Gaussian.\footnote{As described below, we build our model in log flux or magnitudes. As our uncertainties are essentially Gaussian in flux, this poses a challenge at low S/N, where uncertainties can be non-Gaussian and even the log flux values may not be defined for negative flux. We take the log of the simulated noise-free fluxes for this reason.} The parameter covariance matrix is then
\begin{equation} \label{eq:paramcov}
    \left[ J^T \mathrm{(Obs\ Cov)}^{-1} J \right]^{-1}\;.
\end{equation}
The matrix $J_{uv}= \designmatrix$ is called the design matrix or Jacobian matrix. The distance moduli are in our list of parameters (described below), so we can read off the distance-modulus covariance matrix from the appropriate block of the parameter covariance matrix computed by Equation~\ref{eq:paramcov}.

For our calculation all observations (photometric and spectrophotometric) are modeled simultaneously in magnitude (log flux) with the following terms added together for each observation for each SN (the derivatives of this equation with respect to each parameter make up the design/Jacobian matrix for the Fisher calculation):

\newcommand{\Dgray}{\ensuremath{\Delta \mathrm{Gray}_i}\xspace}
\newcommand{\Dfilt}{\ensuremath{\Delta \mathrm{Flt}_{ik}}\xspace}
\newcommand{\Alamb}{\ensuremath{R (\tfrac{\lambda_k}{1 + z_l} ) E(B-V)_i}\xspace}
\newcommand{\DZP}{\ensuremath{\Delta \mathrm{ZP}_{k}}\xspace}
\newcommand{\DWaveBare}{\ensuremath{\Delta \lambda_k}\xspace}
\newcommand{\DWave}{\ensuremath{[\partial m_{ijk}/\partial \lambda] \, \DWaveBare}\xspace}
\newcommand{\DMean}{\ensuremath{\Delta \mathrm{Mean}(\tfrac{\lambda_k}{1 + z_l},\, \tfrac{t_{j} - t^{\mathrm{Max}}_i}{1 + z_l} )}\xspace}
\newcommand{\DCRNLBare}{\ensuremath{\Delta \mathrm{CRNL}}\xspace}
\newcommand{\DCRNL}{\ensuremath{\DCRNLBare \, \log_{10}(\mathrm{CR}_{ijk}/\mathrm{10,000\ e}^{-}/\mathrm{s/pix})}\xspace}
\newcommand{\FundColorBare}{\ensuremath{\Delta \mathrm{Slope}}}
\newcommand{\FundColor}{\ensuremath{\FundColorBare \, (\lambda_k - 1 \mu\mathrm{m})}}

\begin{align} \label{eq:fishermodel}
    m_{ijk} & = \Dgray + \Dfilt + \Alamb  \\
     & + \mu_l + \DZP + \DWave \nonumber \\
     & + \DMean \nonumber \\
     & + \DCRNL \nonumber \\
     & + \FundColor \nonumber \;,
\end{align}
where $i$ runs over SNe, $j$ runs over observations (either photometric or spectroscopic), $k$ runs over filters or wavelengths, and $l$ runs over redshift bins.\footnote{To reduce RAM requirements, we analytically marginalize over all the per-SN parameters, i.e., anything that depends on $i$ \citep{woodbury1950inverting, Amanullah2010}.} These terms are:

\begin{itemize}
    \item \Dgray and \Dfilt are terms for implementing the \scatter model chosen for the calculation (Section~\ref{sec:dispersion}). Empirical SN descriptions (e.g., SALT) show residual scatter around both light-curve fits and the Hubble diagram that is not explained by the uncertainties. In other words, the \Dgray and \Dfilt terms represent unexplained \scatter, preventing well-measured data points from having near infinite weight in the calculation. \Dgray is an achromatic/gray magnitude random offset per SN with a prior around zero appropriate for the gray \scatter of the \scatter model being used (Section~\ref{sec:dispersion}).  \Dfilt is a correlated magnitude offset per filter per SN, also with a prior. This combination of offsets and priors is equivalent to having a per-SN covariance matrix.
    \item \Alamb, host-galaxy extinction per SN assuming a \citet{Cardelli1989} extinction law with $R_V=3.1$. We assume all SNe have the same $R_V$; assuming a distribution of $R_V$ (e.g., \citealt{Mandel2011, Burns2014}) will certainly affect the absolute FoM but should not affect the relative FoM rankings of the surveys.
    \item $\mu_l$, the distance modulus for the supernova's redshift bin (we quantize redshift in bins of 0.05 to enable a smaller distance-modulus covariance matrix of size $\sim 50 \times 50$ without losing meaningful precision). We compute FoM from these covariance matrices, as described in Section~\ref{sec:DETF}.
    \item \DZP, a single correlated zeropoint offset for each filter (correlated across all magnitudes and SNe). We take a prior on these offsets of nominally 5~mmags uncertainty.\footnote{The reader should be somewhat concerned that we use the same value for all filters. For example, the \FRR band will be very challenging to calibrate: it is the bluest filter, so aberrations are largest, two-electron production will happen \citep{Givans2022}, and the undersampling is worst. \FRR is also fractionally widest, so the chromatic corrections will be large. On the other hand, it is in the observer-frame optical, so there is a lot of potential to use well-calibrated optical spectrographs to establish spectrophotometric standards for it \citep{Bohlin2014, Rubin2022SNIFS}. These considerations are simply too detailed for this work or generally other similar survey optimization efforts and will not affect survey optimization unless there are dramatic differences between filters.} For the prism, we assume 100\% correlation between zeropoint calibration of a filter and the prism at the effective wavelength of the filter. We do this by placing a spline through the zeropoint offsets at the effective wavelength of each filter, and applying the spline to the prism.\footnote{Completely correlated calibration is likely a reasonable assumption as there will be many stars observed in both the prism and the imaging. However, the prism may be better calibrated in the end. First, we note the prism has no passband uncertainties, unlike filters. Second, the prism can directly measure many more CALSPEC stars \citep{Bohlin2014} than the imaging due to imaging saturation on bright sources. Third, the prism sees all SNe at essentially the same background-limited count rate, although that does make the prism more sensitive to accurate background subtraction. Again, these points are generally too detailed for this analysis.}
    \item For the filters, correlated per-filter effective-wavelength offsets impact the magnitudes as \DWave, where the derivative is computed with no statistical noise in the simulation. We take a nominal $5\ang$ prior on each of these per-filter offsets, and again assume this prior is the same size for all filters.
    \item The training of the mean SN model \DMean, which is a function of both rest-frame wavelength and phase (rest-frame days with respect to maximum). Here, we use a 2D spline. We assume nine spline nodes uniformly spaced in log wavelength between $3,000 \ang$ and $16,000 \ang$ (c.f., 10 such parameters in \citealt{Astier2011}). We find virtually no impact on FoM of changing the number of nodes, in agreement with \citet{Astier2011}. We exclude data falling outside this wavelength range. The spline has 10 uniformly spaced nodes in time from $-15$ to $+45$~days with respect to maximum light.\footnote{Note that there is no integration of the model over the filter passband in \DMean; it is evaluated exactly at the effective wavelength of each filter (or prism pixel along the dispersion direction). 
    Essentially, we are assuming that the spline model has been convolved (in log wavelength) with the width of a filter. This would be exact if all filters have the same width in log wavelength and the prism measured convolved spectra. In detail, the \FRR is fractionally wider and the \FF is fractionally narrower than the filters, and a prism pixel is much narrower. So this is not a perfect approximation. But the main goal of including the training is to propagate calibration uncertainties, and those are all broadband. Also, as noted above, changing the model spectral resolution has little impact so filter convolution likely will not matter either.} Simulating the training of the mean model is extremely important for propagating calibration uncertainties and it makes a large difference on FoM (see also \citealt{Astier2011}). In contrast, the training of the color law and the light-curve-shape-variation template have little dependence on the calibration, so we do not include these. To eliminate the degeneracy of making the mean SN model fainter and all the distance moduli brighter (or vice versa), we pick the spline node closest to rest-frame $B$-band maximum and fix it to zero. To eliminate the degeneracy of making the mean SN model redder and making all the $E(B-V)_i$ values bluer (or vice versa), we pick the spline node closest to rest-frame $V$-band maximum and fix it to zero as well.
    \item The uncertainty on the count-rate nonlinearity (CRNL). We assume a nominal wavelength-independent uncertainty of 0.000125 magnitudes (0.125~mmag) per dex and assume that this slope uncertainty does not change with magnitude or wavelength. This is a simpler model with much smaller uncertainties than \citet{Deustua2021} who considered flux and wavelength dependence, but the results of ground-based testing are encouraging so far. We are also encouraged that a precise measurement should be possible on orbit using the Relative Calibration System. We include all sources of count-rate in our computation, not just from the SNe and take 10,000~e$^-$/pixel/second as the reference count rate. The ground-based (Rubin Observatory) CCD data are assumed to have no CRNL.
    \item A fundamental color calibration slope uncertainty in wavelength \FundColor, representing the limit of knowledge of the spectrophotometric standards as a function of relative wavelength (c.f., \citealt{Bohlin2020}, Figure~4). SNe~Ia are calibrated to nearby SNe, so only the relative calibration as a function of wavelength matters here, thus the 1~$\mu$m is arbitrary.\footnote{This is not true when calibrating to theoretical models to obtain the Hubble constant \citep{Hoeflich1996, Vogl2024}, which is beyond the scope of this work.} We take a nominal prior of $\pm 7$~mmags/$\mu$m on $\Delta$Slope.
\end{itemize}

We do not include weak-lensing dispersion or intergalactic extinction, as we assume these will be regressed out (e.g., for lensing, \citealt{Shah2024}). We also assume Milky-Way extinction is small ($E(B-V) \sim 0.01$) for the survey fields and neglect it as it cannot affect the survey ranking.

In most analyses, SN selection cuts are functions of wavelength, light-curve sampling, S/N, and parameters/uncertainties from the light-curve fit. Here, we approximate the selection of SNe by applying a cut on the quadrature sum of S/N as motivated in Appendix~\ref{sec:selectioncuts}. For shorthand, we call this ``\SNRSUM'' and apply it variously to imaging and spectroscopic S/N over wavelengths/filters/epochs:
\begin{equation} \label{eq:SNRSum}
    \mathrm{SNR\ Sum} \equiv \sqrt{\sum [\mathrm{S/N}]^2}\;.
\end{equation}
We require that \SNRSUM, sum across all imaging filters and epochs is $>$~40 (this is roughly equivalent to having \SNRSUM $>20$ across all epochs in each of four filters so that SN color can be measured to $\lesssim 0.05$ magnitudes, and is also motivated in Appendix~\ref{sec:selectioncuts}). 
As the cadence and depths are assumed to be very stable (e.g., no weather, no moonlight, and little PSF variation), SN selection based on total S/N over the light curve (i.e., \SNRSUM) is a good approximation.\footnote{The SNe from the Rubin DDF will have weather and seeing variation, but we assume a conservative volume-limited subset at high S/N, so we do not worry about them here.}

\subsection{\Scatter Model} \label{sec:dispersion}

Section~\ref{sec:Simulations} describes the sources of noise in the simulations, but this is not enough to compute an accurate FoM value; one must also consider the \scatter of the SNe themselves (this is sometimes referred to as ``intrinsic'' or ``unexplained'' \scatter). We consider two dispersion models to investigate the sensitivity of the optimization to this assumption: ``NIR lower'' and ``Twins lower.'' The NIR-lower model assumes that rest-frame NIR data have lower scatter and should have more weight; conversely the Twins-lower model assumes that standardization is improved when the SN is observed with the prism. These are plausible bookends of the true scatter model which presumably will have elements of both of these models. An important note is that both \scatter models fit all the data for each SN (in the rest-frame/phase range of the model), including both rest-frame optical/NIR and imaging/prism data. They simply weight the data differently and assume different forms for the unexplained \scatter. We note the two \scatter models are not scaled to give the same FoM on average and the absolute values of FoM are not very meaningful; therefore the two \scatter models' FoM values cannot be directly compared. However, within each model, the different survey options can be ranked, and the overall goal is to optimize over both of these possible models together. Section~\ref{sec:Results} compares FoM values using both models.

\subsubsection{NIR-Lower \Scatter Model}

\newcommand{\NIRLowerSentence}{Strategies with more filters average down the per-filter correlations and strategies covering a larger wavelength range enable better measurements of extinction.\xspace}

The NIR-lower \scatter model assumes 0.08~mag gray \scatter, 0.04~mag correlated-in-phase \scatter for filters with effective wavelengths $< 9,000\ang$, and 0.02~mag correlated \scatter in the NIR (c.f., \citealt{Pierel2022} Figure~8).\footnote{Other analyses have made other assumptions. For example, \citet{Astier2011} assumed 0.025~magnitudes. In addition to the NIR-lower \scatter model, we tested a model that had equal per-band \scatter as a function of wavelength and found it similar so it is not included here.} The phase-and-filter-correlated uncertainty is introduced to the model in Equation~\ref{eq:fishermodel} with the \Dgray term. The per-filter correlated scatter is introduced with the \Dfilt term. The prism is assumed to have no such \Dfilt term as much of the color \scatter seems to come from spectral-feature variation that can be measured with spectrophotometry and regressed out \citep{Chotard2011}. Table~\ref{tab:NIRLower} shows the implied distance-modulus dispersion in the limit of infinite S/N in each filter. \NIRLowerSentence These values roughly match, e.g., \citet{Mandel2022}.

\begin{deluxetable*}{lcc}
\label{tab:NIRLower}
\tablecaption{Implied distance-modulus dispersions for the NIR-lower model in the limit of infinite S/N in each filter. \NIRLowerSentence}
    \tablehead{\colhead{Rest-Frame Filters}	&	\colhead{Distance Uncertainty} &	\colhead{Distance Uncertainty Including Gray Dispersion}}
\startdata
$BV$ & 0.197 & 0.212 \\
$BVR$ & 0.109 & 0.135 \\
$UBVR$ & 0.081 & 0.114 \\
$UBVRI$ & 0.055 & 0.097 \\
$BVRIY$ & 0.030 & 0.085 \\
$BVRIYJ$ & 0.021 & 0.083
\enddata
\end{deluxetable*}

\subsubsection{Twins-Lower \Scatter Model}

\newcommand{\KBoone}{K. Boone, private communication for the Perlmutter SN Roman Science Investigation Team\xspace}

\begin{figure}[htbp]
    \centering
    \includegraphics[width=0.49\textwidth]{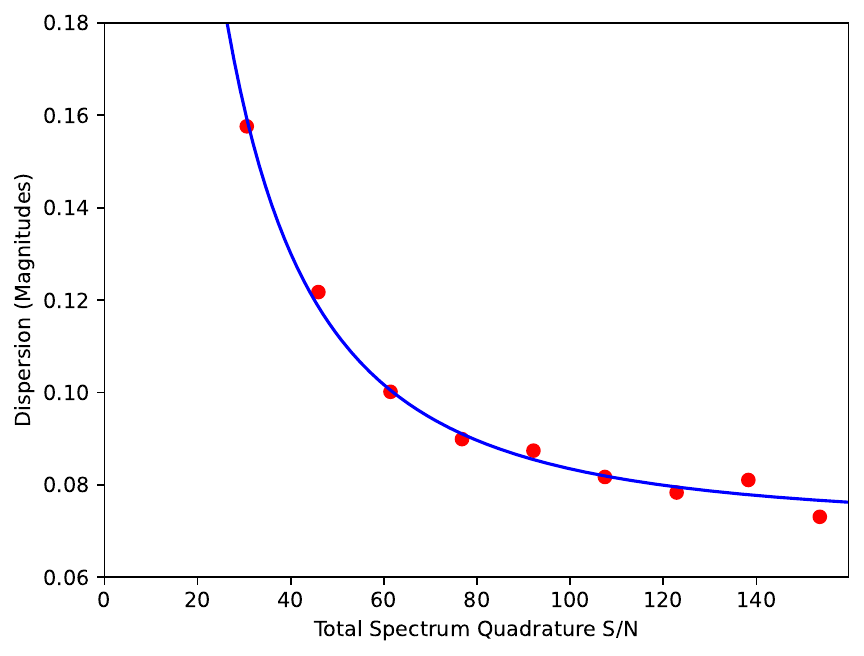}\\
    \includegraphics[width=0.49\textwidth]{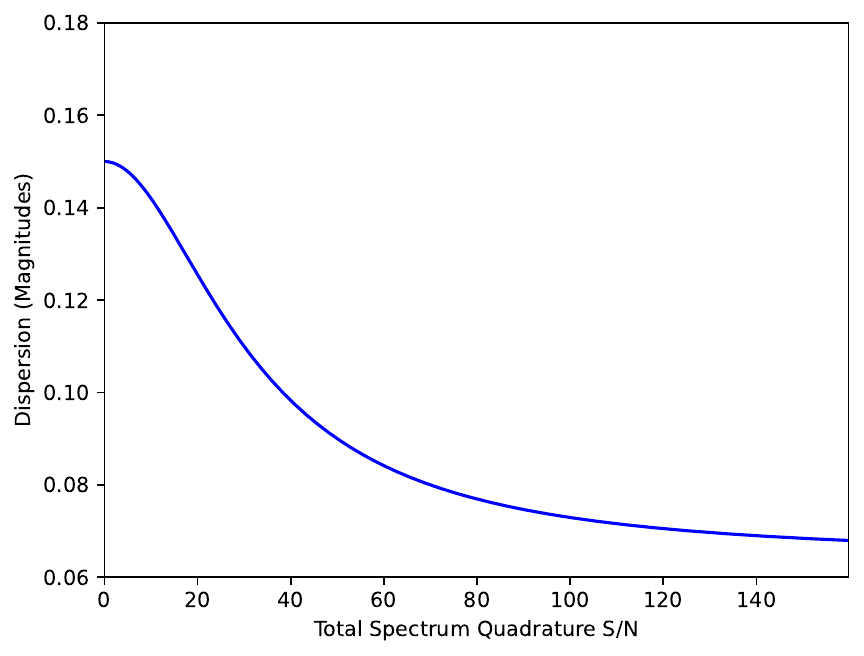}
    \caption{{\bf Top panel:} Twins analysis dispersion \citep{Fakhouri2015} as a function of degraded overall quadrature S/N for a spectrum (\KBoone) which we fit with Equation~\ref{eq:twins}. {\bf Bottom panel:} The same curve combined with a weighted average of an assumed 0.15~magnitude floor which applies when the spectroscopic observations are missing or noisy (Equation~\ref{eq:twinsfloor}). We refer to this curve as the ``Twins lower'' \scatter model.
    \label{fig:twins}}
\end{figure}

The Twins-lower \scatter model is based on the observed twin-SN dispersion (\citealt{Fakhouri2015}) as a function of overall spectral S/N (\KBoone and reproduced in Figure~\ref{fig:twins}). We fit these points with a part that varies with S/N and a constant floor:
\begin{equation} \label{eq:twins}
    \mathrm{RMS\ from\ Twins} = \sqrt{0.0712^2 + \frac{18.96}{\mathrm{S/N}^2}}\;;
\end{equation}
the floor value roughly matches \citet{Boone2021} and \citet{Stein2022}. We sum S/N in quadrature over all phases and wavelengths (\SNRSUM, Equation~\ref{eq:SNRSum}).\footnote{The twin-SN dispersion as a function of S/N was computed assuming a single spectrum at maximum although we apply it to the S/N of a time series; see \citet{Rubin2022} Figure~5 for how time-series S/N is distributed around maximum light. Presumably the right approach is to build a full optical--NIR spectral-time-series model that can fit both spectroscopy and imaging and has $\sim$~three components of SN SED variation (e.g., \citealt{Leget2020, Stein2022}). With both cadenced imaging and spectroscopy, \RomanST should be able to train such a model.} Then we take a weighted average against 0.15~mag dispersion from the imaging (which is the fallback when spectral observations is missing or very low S/N), showing the curve in the bottom panel of Figure~\ref{fig:twins}:
\begin{equation} \label{eq:twinsfloor}
    \mathrm{RMS} = \sqrt{0.0643^2 + \frac{12.63}{26.22^2 + \mathrm{S/N}^2}}\;.
\end{equation}
The leading term at high S/N is smaller because Equation~\ref{eq:twinsfloor} includes imaging as well. This phase-and-filter-correlated uncertainty is introduced to the model in Equation~\ref{eq:fishermodel} with the \Dgray term.

\subsection{Computing Figures of Merit} \label{sec:DETF}
After computing the distance-modulus covariance matrix, we compute a $w_0$-$w_a$ covariance matrix assuming a flat universe and a 0.26\% constraint on the shift parameter from Planck \citep{Efstathiou1999, PlanckCollaboration2020}:
\begin{equation}
    R \equiv \sqrt{\Omega_m} \int_{z^{\prime} = 0}^{1090} \frac{dz^{\prime}}{\sqrt{E(z^{\prime})}}
\end{equation}
with $E(z)$ equal to
\begin{equation}
        \Omega_m (1+z)^3 + (1-\Omega_m) e^{-\frac{3 w_a z}{1 + z}} (1 + z)^{3 \left(\frac{w_a}{1 + z_p} + w_p+1\right)}
\end{equation} 
(both for a flat universe). Here, $z_p$ is the pivot redshift, $w_p$ is the dark-energy equation-of-state parameter at $z_p$, and $w_a$ is its variation with scale factor. We verify that selecting a pivot redshift of 0.3 (roughly the actual pivot redshift value) or 0 (in which case $w_p$ is equal to $w_0$) does not matter for the FoM value computed. One could directly compute the $w_0$-$w_a$ covariance matrix without going through the distance-modulus covariance matrix, but computing the distance-modulus covariance matrix for each survey enables future work to consider alternatives to $w_0$-$w_a$.\footnote{Our distance-modulus covariance matrices are available at \url{https://github.com/rubind/roman_fisher_results}.} With the covariance matrix computed, $(\mathrm{Det(Cov}(w_0,\, w_a))^{-1/2}$ gives the \citet{Wang2008} variant adopted by the \RomanST Science Definition Team of the Dark Energy Task Force Figure of Merit (DETF FoM, \citealt{Albrecht2006}).\footnote{The true DETF FoM is defined as the inverse area of the 95\% confidence ellipse, but this is only off from our definition by a constant and we follow the \RomanST Science Definition Team convention here. The true DETF FoM also fits for curvature but again we follow the SDT and assume a flat universe.}

We take $w_0=-1$, $w_a=0$ as our fiducial cosmology. However, given the possible indications of time-evolving dark energy \citep{Rubin2023UNITY, DESCollaboration2024, DESICollaboration2024}, it is worth considering alternatives to $\Lambda$CDM. We find a significant cosmology dependence to the FoM, which is plausible, as when there is more dark energy, it is easier to measure its properties. For example, considering Roman-only (no Rubin DDF SNe~Ia) with the NIR-lower \scatter model, the median FoM drop is 13\% for $\Omega_m=0.30$ compared to $\Omega_m=0.28$ (which has more dark energy). However, the scatter survey-to-survey is small, so the ranking of the surveys does not change. Considering a time-evolving dark energy model from \citet{Rubin2023UNITY} ($\Omega_m=0.329$,~$w_0=-0.744$,~$w_a=-0.79$), the FoM drops by a further 9\% (median) with again small scatter.\footnote{The RMS scatter around a linear regression line between FoM computed with $\Lambda$CDM and FoM computed with this time-evolving $w_0$-$w_a$ model is 0.7\%.} We conclude that moderate perturbations in the choice of fiducial cosmological model are unlikely to change the choice of optimum HLTDS.

\newcommand{\wpwaconclusions}{The uncertainty on $w(z=0.3)$ strongly correlates with FoM and the uncertainty on $w_a$ has almost 100\% correlation with FoM. We conclude that the weighting of these uncertainties is unlikely to have a significant impact on the choice of optimum HLTDS.\xspace}

\begin{figure}
    \centering
    \includegraphics[width=0.4\textwidth]{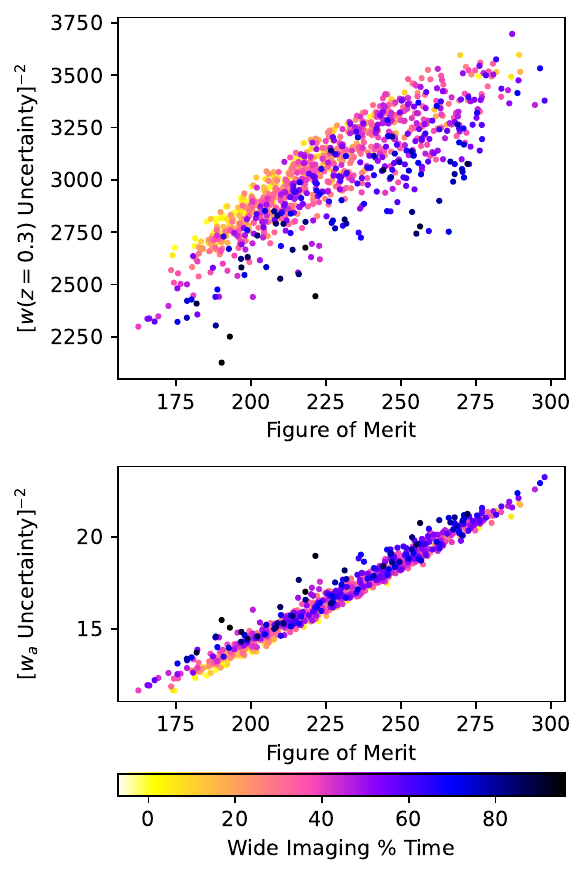}
    \caption{{\bf Top panel:} uncertainty on $[w(z=0.3)]^{-2}$ as a function of Figure of Merit (which is also an inverse squared uncertainty) color coded by the fraction of the time in Wide imaging. {\bf Bottom panel:} a similar plot for the uncertainty on $w_a^{-2}$. \wpwaconclusions
    \label{fig:wpwaFoM}}
\end{figure}

This work only considers the combination of SNe and CMB. It is thus also worth decomposing the Figure of Merit into its constituents of the uncertainty on $w_{\mathrm{pivot}}$ (roughly $w(z=0.3)$) and $w_a$ as different data combinations will have different parameter degeneracies that weight these uncertainties differently. Figure~\ref{fig:wpwaFoM} shows scatter plots of these uncertainties$^{-2}$ versus FoM for \RomanST alone (evaluated with the NIR-lower \scatter model) for our \numsurveys surveys. \wpwaconclusions

The \RomanST Science Definition Team made similar assumptions \citep{Spergel2015}, but with a 0.2\% constraint from CMB S4, which now may be delayed. We find 9\% lower FoM values (9\% with a scatter of 0.8\%) when using a 0.26\% shift-parameter constraint as opposed to 0.2\%.

\section{Results} \label{sec:Results}

Our first and most important result is that there is real dynamic range in our relative FoM values; some surveys really are much better than others in terms of constraining power for $w_0$-$w_a$. This is not a trivial point! All of the surveys simulated are reasonable (a good selection of optical elements, cadences suitable for measuring SNe~Ia, exposure times chosen to measure $z\sim 1$ SNe~Ia) and all use the same amount of total observatory time (0.5~years).

\newcommand{\morescatterfindings}{The left panels show significant scatter between the two models when Rubin DDF SNe~Ia are not included, with the best surveys according to the Twins-lower \scatter model having more prism and less Wide-imaging area (red histograms) and the NIR-lower \scatter model favoring the opposite (blue histograms). With Rubin DDF SNe~Ia, the pressure to cover the maximum possible area is relieved so both models correlate much more strongly and the histograms show much more overlap.\xspace}

\begin{figure*}[htbp]
    \centering
    \includegraphics[width=0.9 \textwidth]{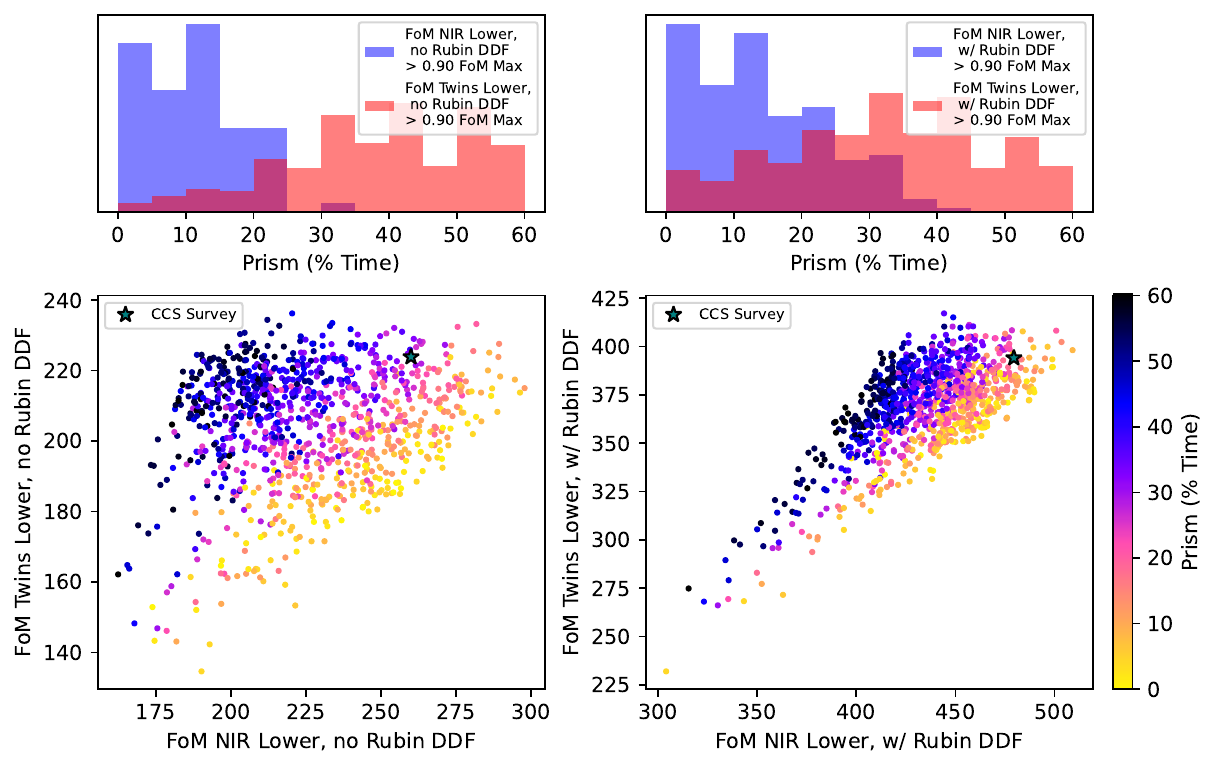}
    \includegraphics[width=0.9 \textwidth]{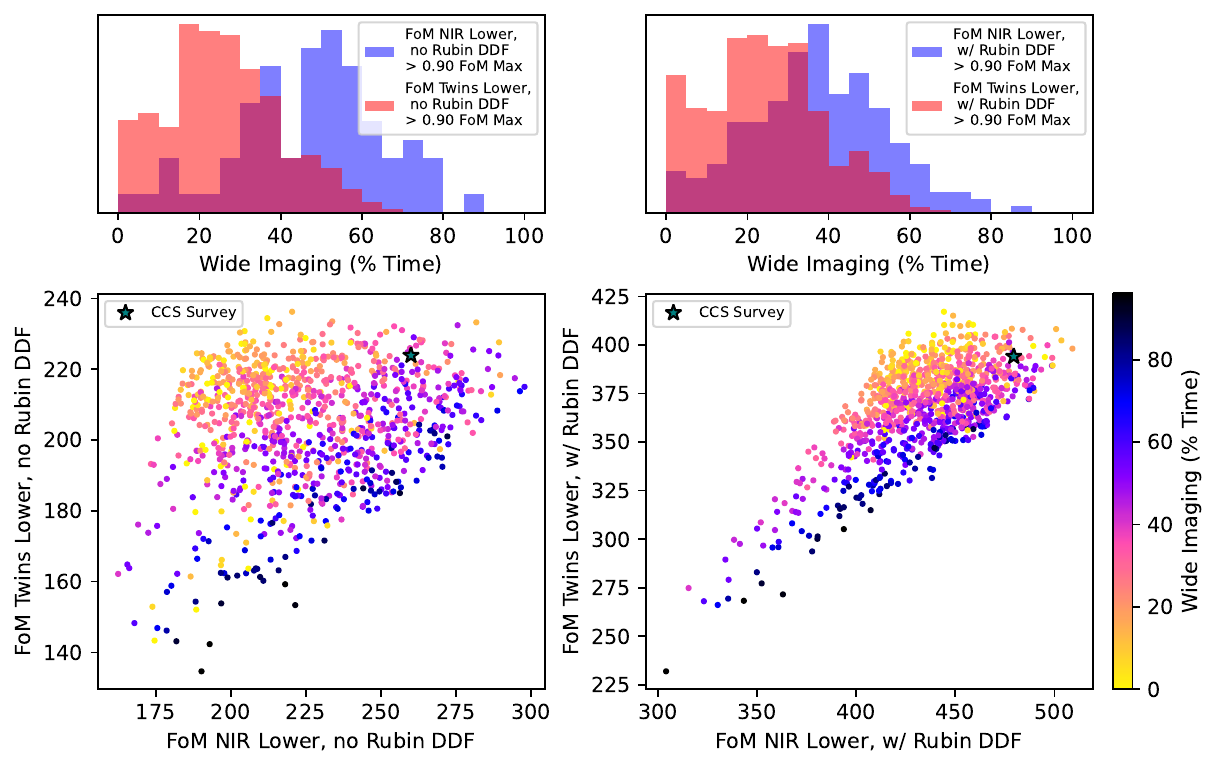}
    \vspace{-0.15 in}

    \caption{Scatter plots comparing FoM values (including systematics) from our two \scatter models (NIR lower on the $x$ axis, and Twins lower on the $y$ axis) both without Rubin DDF SNe~Ia ({\bf left panels}) and with Rubin DDF SNe~Ia ({\bf right panels}). Again, we remind the reader that these two \scatter models are not scaled to give the same FoM on average and are only designed to help rank the surveys. The color coding is by percentage of time going to the prism in the {\bf top panels} and by percentage of time going to the Wide-imaging tier in the {\bf bottom panels}; the remainder of the time goes to the Deep-imaging tier. \morescatterfindings The \CCSsurvey (discussed in Section~\ref{sec:CCSSurvey} is marked with a star and is near the top of the rankings when Rubin DDF SNe are considered.
    \label{fig:scatter}}
\end{figure*}

Figure~\ref{fig:scatter} compares our two \scatter models, with (right panels) and without (left panels) assuming Rubin DDF SNe~Ia. One immediate finding is that the FoM gets much larger when Rubin DDF SNe~Ia are included. Although this is certainly true, the exact magnitude may be overstated here, as the nearby SNe are also assumed to be taken by Rubin (also in $griz$), or another instrument that is perfectly inter-calibrated with Rubin. The histograms show surveys with assessed FoM values $>$~90\% of the maximum FoM. The color bar on the top scatter plots and the top histograms show the percentage of time used by the prism; the bottom shows the percentage of time used in the Wide-imaging tier. \morescatterfindings

\begin{figure}[htbp]
    \centering
    \includegraphics[width=0.49 \textwidth]{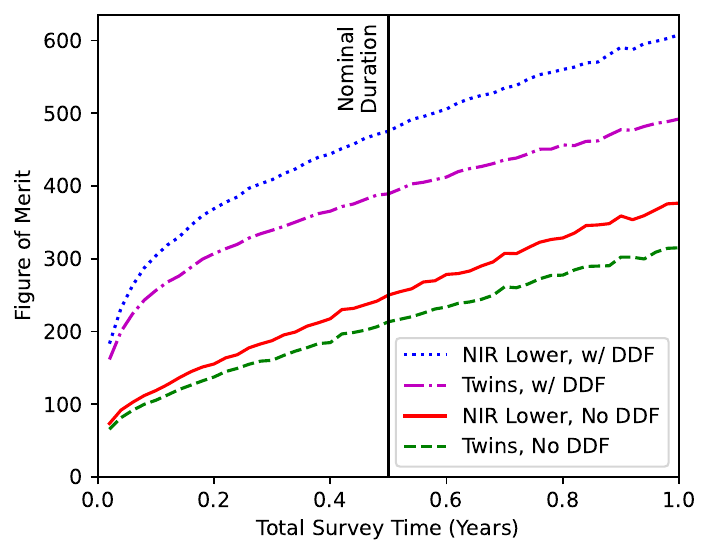}
    \caption{FoM vs total \RomanST survey time for both of our \scatter models and with and without Rubin DDF (all computed for a survey with filters and exposure times similar to the \CCSsurvey, but scaled in area). These results suggest that the HLTDS can still gain in constraining power past its nominal 0.5 year allotment. As a reminder, the goal for this work is to rank order proposed surveys rather than compute the absolute FoM, which may require a more detailed model for both statistical and systematic uncertainty. The two \scatter models are also not scaled to give the same FoM on average (there are significant uncertainties in the size of the \scatter so the FoM values should be compared within a given scatter model and not between them). Also, the 800 low-$z$ SNe are assumed to be on the Rubin Observatory magnitude system; if this is not achieved, this may overstate the relative contribution of the Rubin DDF SNe.}
    \label{fig:FoMTime}
\end{figure}

Figure~\ref{fig:FoMTime} examines the impact of running the HLTDS shorter or longer than the nominal 0.5 years. This figure shows all four possibilities: both \scatter models, with and without Rubin DDF. The FoM curves continue to increase all the way past 1 year.

\newcommand{\withwithouttraining}{with and without simultaneously training the mean-SN model so that its uncertainties can be propagated.\xspace}
\newcommand{\trainingmatters}{The FoM values are lower when including training, of course, but the FoM also has much more sensitivity to the systematic uncertainties when training uncertainties are propagated, so it is very important to do this. This figure shows both the FoM vs. the prior values (dotted lines) and vs. the posterior values (solid lines). The improvement in the posterior is due to self calibration \citep{Kim2006}. We note that the exact amount of self calibration will depend on the accuracy of the model assumptions.\xspace}

\begin{figure*}[htbp]
    \centering
    \includegraphics[width=0.9 \textwidth]{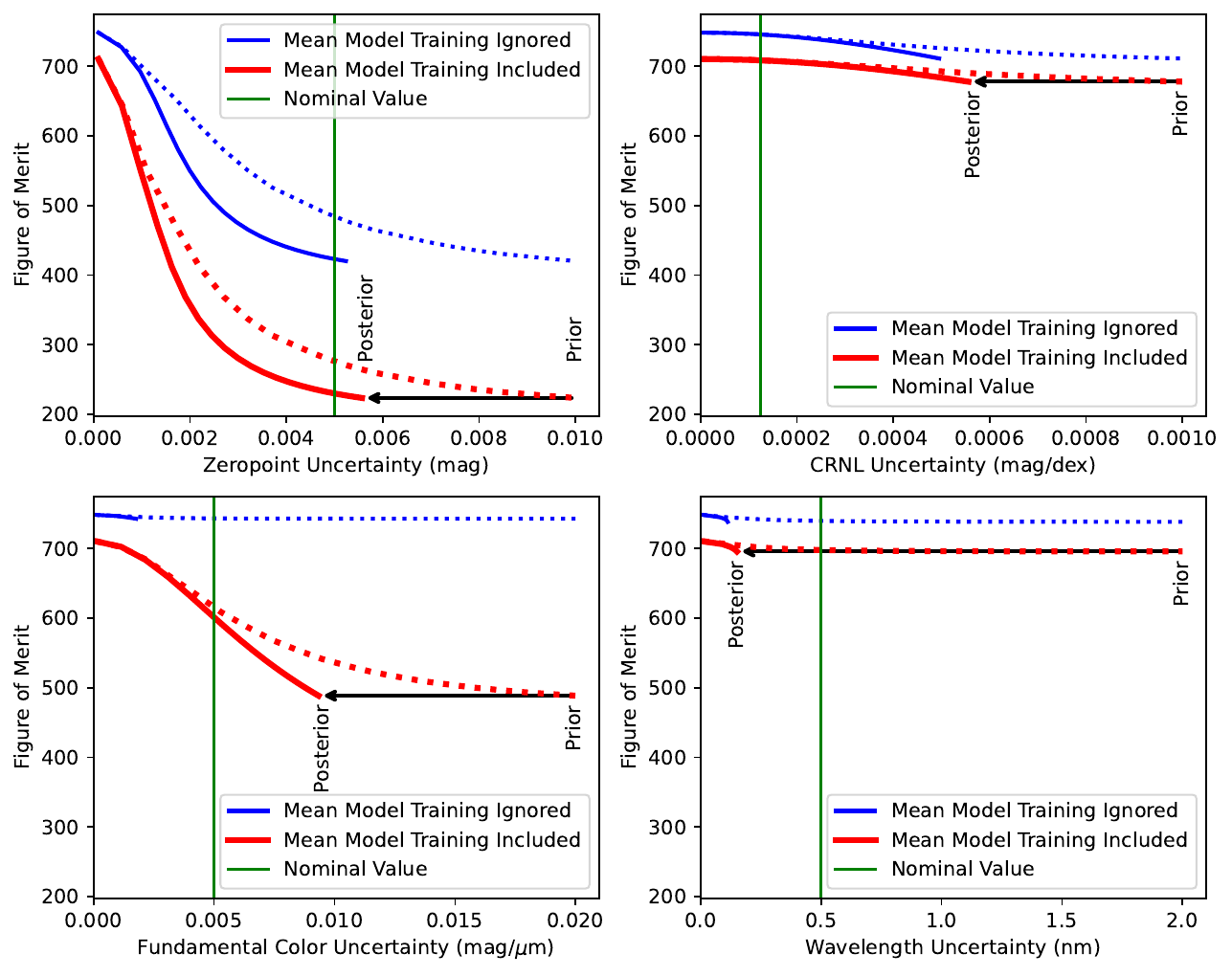}
    \caption{Computed FoM for the \CCSsurvey as a function of assumed systematic uncertainties using the NIR-lower \scatter model, \withwithouttraining The {\bf upper-left panel} shows FoM vs. the independent per-filter zeropoint uncertainty (the size of the priors on \DZP), the {\bf upper-right panel} shows the CRNL uncertainty (the size of the prior on \DCRNLBare), the {\bf lower-left panel} shows the fundamental (or ``absolute'') color-slope uncertainty (the size of the prior on \FundColorBare), and the {\bf lower-right panel} shows the independent per-filter effective wavelength uncertainty (the size of the priors on \DWaveBare). \trainingmatters  Self calibration is quite a large effect for effective-wavelength uncertainties, in that the posterior uncertainties are much smaller ($\sim 0.1$~nm) than the prior uncertainties which range up to 2~nm.
    \label{fig:FoMvsUnc}}
\end{figure*}

Figure~\ref{fig:FoMvsUnc} shows the impact of varying the assumed calibration uncertainties on the \CCSsurvey FoM (assuming the NIR-lower \scatter model), \withwithouttraining \trainingmatters We find that the independent per-filter zeropoint uncertainties have the largest impact on FoM and the fundamental color slope the second largest, although one should really think of these ``short-range-in-wavelength'' and ``long-range-in-wavelength'' calibration uncertainties as two components of the whole. Having a linear slope in wavelength for the fundamental color uncertainties is likely not as bad as a more flexible parameterization (e.g., a quadratic or even per-filter uncertainties) because a slope in wavelength simply moves all SNe redder and bluer by a similar amount as a function of redshift and thus partially cancels out. The CRNL uncertainties have a much smaller impact on FoM than were computed by \citet{Deustua2021}; this is due both to the more rigid parameterization here (a constant-in-wavelength-and-flux power law) and the assumed smaller size. Finally, the passband effective-wavelength uncertainties are self calibrated to $\sim 0.1$\,nm ($\sim 1\ang$) for the typical filter and so have little effect. Furthermore, we note this calculation ignores the impact of calibration with sources of non-SN spectral energy distributions like quasars, emission-line galaxies, stars, and brown dwarfs which could plausibly improve the constraints beyond $1\ang$.

\newcommand{\otherparstext}{As expected from observing overheads and read noise \citep{Rubin2023Cadence}, slower Wide-imaging cadences increase the FoM, with a smaller effect from making the Deep-imaging cadence slower. Introducing \FF into the Wide-imaging tier reduces the area covered and does not seem to be worth it according to our \scatter models. There is mild disagreement about what redshift should be targeted by the Deep-imaging tier, with the NIR-lower \scatter model and no Rubin DDF SNe preferring more area at less depth, the NIR-lower \scatter model with Rubin DDF relatively insensitive to targeted redshift, and the Twins-lower \scatter model preferring a deeper Deep-imaging tier, presumably to get better imaging S/N on the SNe observed with the prism.\xspace}

\begin{figure*}[htbp]
    \centering
    \includegraphics[width=0.9 \textwidth]{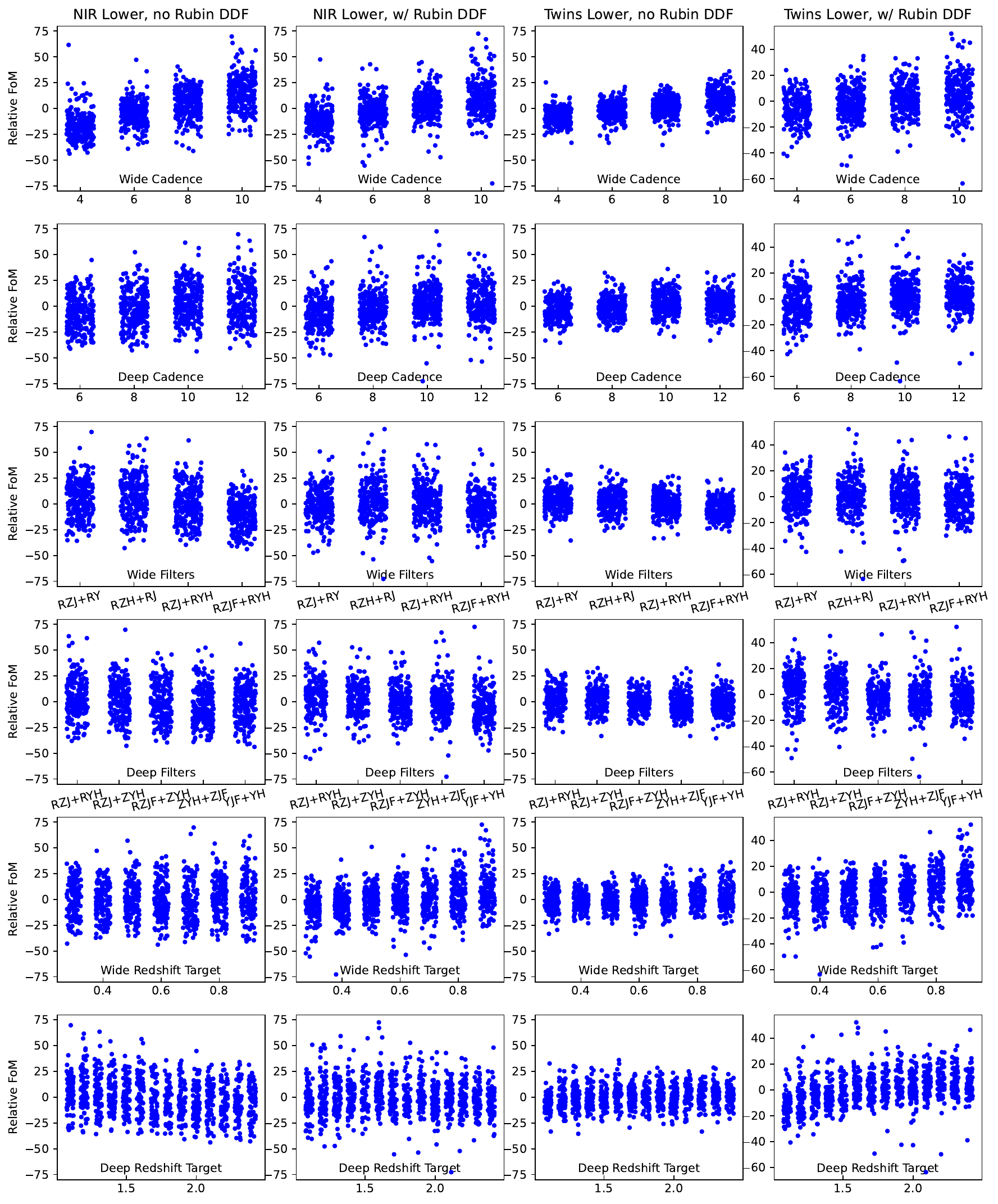}
    \caption{Relative FoM (including systematics) as a function of survey choices that matter less than the relative allocations of time. We regress out the relative allocation of time in the Wide-imaging tier, Deep-imaging tier, and prism to better highlight the differences for the specific choices shown here. Each {\bf column} shows a different \scatter model or inclusion/exclusion of Rubin DDF SNe. Each {\bf row} shows a different survey parameter: cadence in the Wide-imaging tier, cadence in the Deep-imaging tier, Wide-tier filters, Deep-tier filters, Wide-tier redshift target, and Deep-tier redshift target (top to bottom). \otherparstext
    \label{fig:OtherParams}}
\end{figure*}

Finally, Figure~\ref{fig:OtherParams} shows the impact of survey choices other than the relative time allocations on FoM for both of our \scatter models and with and without Rubin DDF SNe. To help highlight the impact of these choices, we regress out the dependence of FoM on the fraction of time going to the Deep-imaging tier and Wide-imaging tier with a cubic polynomial and show the residuals in Figure~\ref{fig:OtherParams}. \otherparstext

\subsection{Simulations of the \CCSsurvey} \label{sec:CCSSurvey}

In this section, we show specific simulations of the \CCSsurvey (\url{https://asd.gsfc.nasa.gov/roman/comm\_forum/forum\_17/Core\_Community\_Survey\_Reports-rev03-compressed.pdf}). We approximate the survey as having a homogeneous cadence for two years, although the actual CCS recommendation covers slightly less area but extends the time with some pre- and post-two-year visits. Compared to the \citet{Rose2021} \ReferenceSurvey with 25\% prism, the \CCSsurvey has roughly the same Wide-imaging area, double the Deep area, a fifth band in both tiers (\FH added to the Wide-imaging tier and \FZ added to the Deep-imaging tier), and 22\% prism instead of 25\%. In detail, both tiers have an \interlaced cadence of 10 days, with the prism and half the filters every five days. Table~\ref{tab:CCSsurvey} highlights the relevant parameters. We assume 146 visits for the filters with a 5-day cadence (two years nominal survey length), times the 0.91 fill factor described above. We assume 146 visits for the prism, as the exposures are long enough to dither over the chip gaps (thus the effective prism exposure time is reduced by 0.91, but the number of visits is 146). The Wide-tier filters are \FRR (5-day cadence), \FZ, \FY, \FJ, \FH (ten-day cadence) and the Deep-tier filters are \FZ (five-day cadence), \FY, \FJ, \FH, and \FF. We note that the areas covered are not integer multiples of the Roman FoV; \citet{Rubin2022} Figure~4 shows how pointings may be overlapped to approximate desired field shapes.

\begin{deluxetable*}{lrrrrrrr}
\label{tab:CCSsurvey}
\tablecaption{Parameters for the baseline \CCSsurvey. Quoted $5\sigma$ depths assume isolated point sources with no galaxy background. These survey areas differ from the CCS report as we approximate the survey as homogeneous in time over two years while the actual \CCSsurvey has a pre- and post-two-year extension and thus covers a bit less area. The 0.91 is the approximate fill factor of the focal plane; SNe falling in chip gaps have missed epochs visible in Figure~\ref{fig:MedianLCs}. We note the CCS recommendation is to dither between filters, so SNe will not have missing epochs in more than one filter.}
    \tablehead{\colhead{Parameter}	&	\colhead{Prism} &	\colhead{\FRR}	&	\colhead{\FZ}	&	\colhead{\FY}	&	\colhead{\FJ}	&	\colhead{\FH}	&	\colhead{\FF}}
\startdata
\multicolumn{8}{c}{Wide-Imaging Tier, 23.27 Deg$^2$}															\\
Exposure Time per Visit (sec)	&		&	60	&	86	&	96	&	153	&	293	&		\\
Single-Visit Limiting Magnitude (AB, 5$\sigma$)	&		&	25.36	&	25.45	&	25.55	&	26.00	&	26.49	&		\\
Mean Visits in Two Years	&		&	$146 \cdot 0.91$	&	$73 \cdot 0.91$	&	$73 \cdot 0.91$	&	$73 \cdot 0.91$	&	$73 \cdot 0.91$	&		\\
Co-Added Limiting Magnitude (AB, 5$\sigma$)	&		&	28.01	&	27.73	&	27.83	&	28.28	&	28.77	&		\\
\hline															
\multicolumn{8}{c}{Deep-Imaging Tier, 8.23 Deg$^2$}															\\
Exposure Time per Visit (sec)	&		&		&	194	&	294	&	307	&	419	&	1637	\\
Single-Visit Limiting Magnitude (AB, 5$\sigma$)	&		&		&	26.38	&	26.72	&	26.67	&	26.78	&	27.11	\\
Mean Visits in Two Years	&		&		&	$146 \cdot 0.91$	&	$73 \cdot 0.91$	&	$73 \cdot 0.91$	&	$73 \cdot 0.91$	&	$73 \cdot 0.91$	\\
Co-Added Limiting Magnitude (AB, 5$\sigma$)	&		&		&	29.03	&	29.00	&	28.95	&	29.06	&	29.39	\\
\hline															
\multicolumn{8}{c}{Wide Prism Tier, 5.01 Deg$^2$, Inside Deep-Imaging Tier}															\\
Exposure Time per Visit (sec)	&	$900 \cdot 0.91$	&		&		&		&		&		&		\\
Mean Visits in Two Years	&	146	&		&		&		&		&		&		\\
\hline															
\multicolumn{8}{c}{Deep Prism Tier, 0.72 Deg$^2$, Inside Deep-Imaging Tier}															\\
Exposure Time per Visit (sec)	&	$3600 \cdot 0.91$	&		&		&		&		&		&		\\
Mean Visits in Two Years	&	146	&		&		&		&		&		&		
\enddata
\end{deluxetable*}

\newcommand{\secondhalfsentence}{sum of S/N in imaging (\SNRSUM) for the \CCSsurvey also showing the 800 assumed nearby SNe and $z<0.5$ Rubin Observatory DDF SNe (for when these SNe are included in the forecast).\xspace}

\newcommand{\lightcurvescomments}{The median S/N light curve is shown for each redshift, so these are representative. Some missing points are visible due to chip gaps, but in general, the light curves are high quality within the redshift range targeted by each tier.\xspace}

\begin{figure}[htbp]
    \centering
    \includegraphics[height=0.85\textheight]{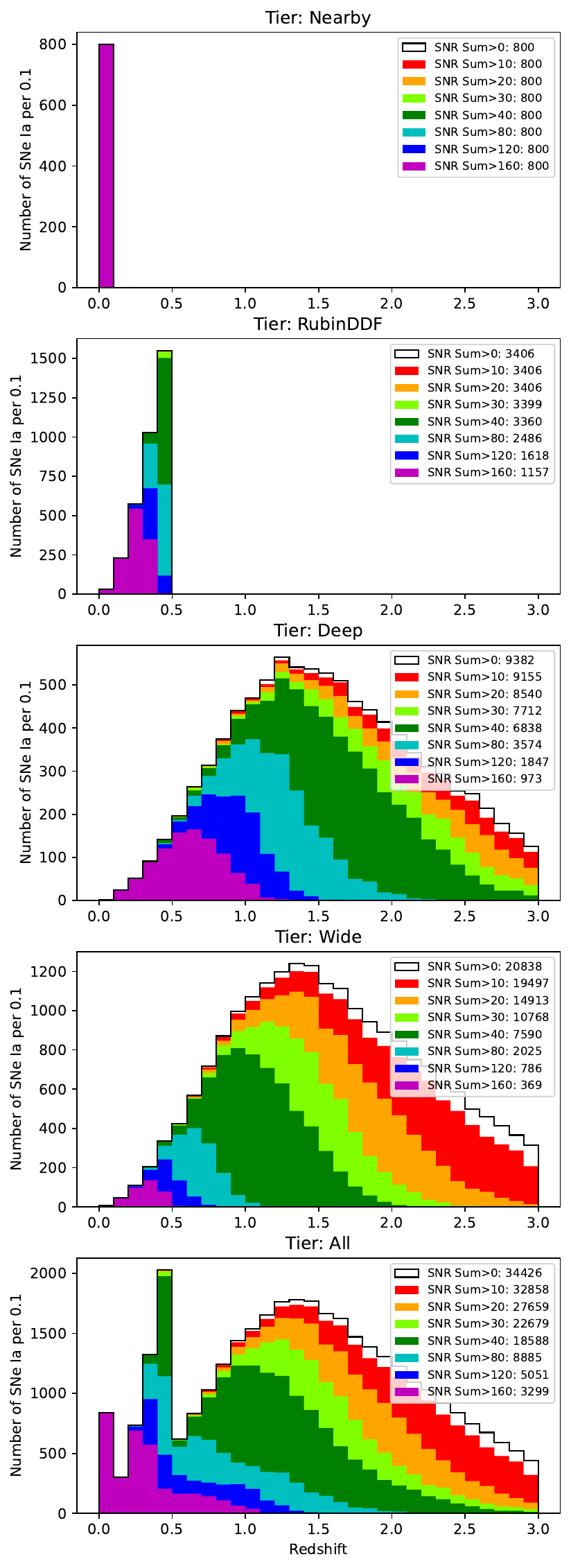}
    \caption{Quadrature \secondhalfsentence The colors show different S/N thresholds. The outlined black histogram shows all SNe~Ia generated, even those at very low S/N.  Our Fisher-matrix analysis approximates real selection cuts with the requirement that total imaging S/N be at least 40 (dark green SNe). Each {\bf panel} shows a different tier and the {\bf bottom panel} shows the combination of all tiers.}
    \label{fig:ImagingSNR}
\end{figure}

\begin{figure*}[htbp]
    \centering
    \includegraphics[height=0.92\textheight]{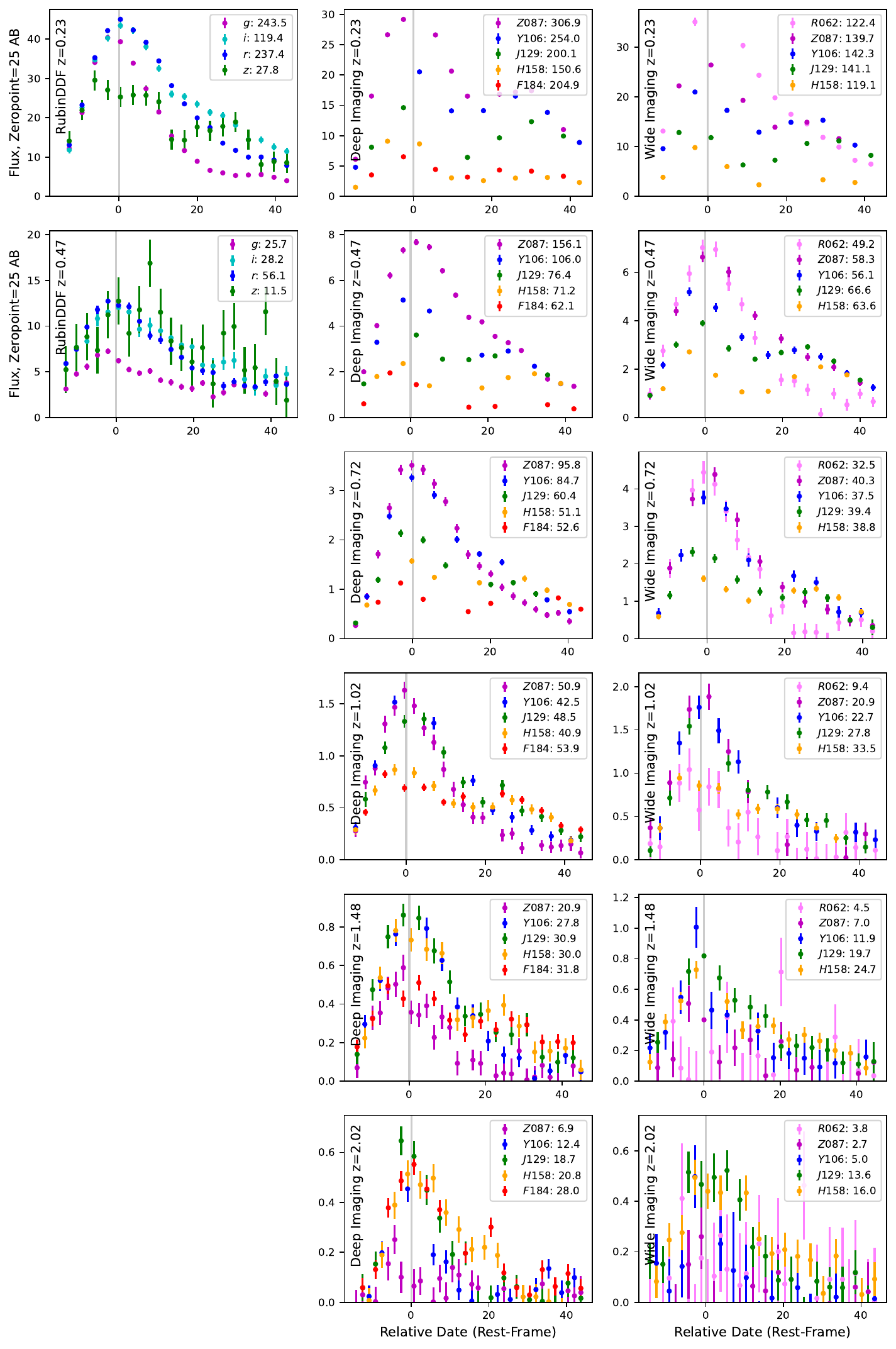}
    \caption{Typical light curves as a function of redshift. The {\bf left column} shows light curves from the Rubin DDF, the {\bf middle column} shows light curves from the Deep-imaging tier, and the {\bf right column} shows light curves from the Wide tier. Each legend shows the total quadrature sum of S/N for each band. We note that in the bluer filter chosen for faster cadence (\FRR in the Wide tier and \FZ in the Deep-imaging tier) the lower S/N per visit roughly balances the additional visits as expected from Equation~\ref{eq:exptime}. \lightcurvescomments}
    \label{fig:MedianLCs}
\end{figure*}

\begin{figure}[htbp]
    \centering
    \includegraphics[width=0.49\textwidth]{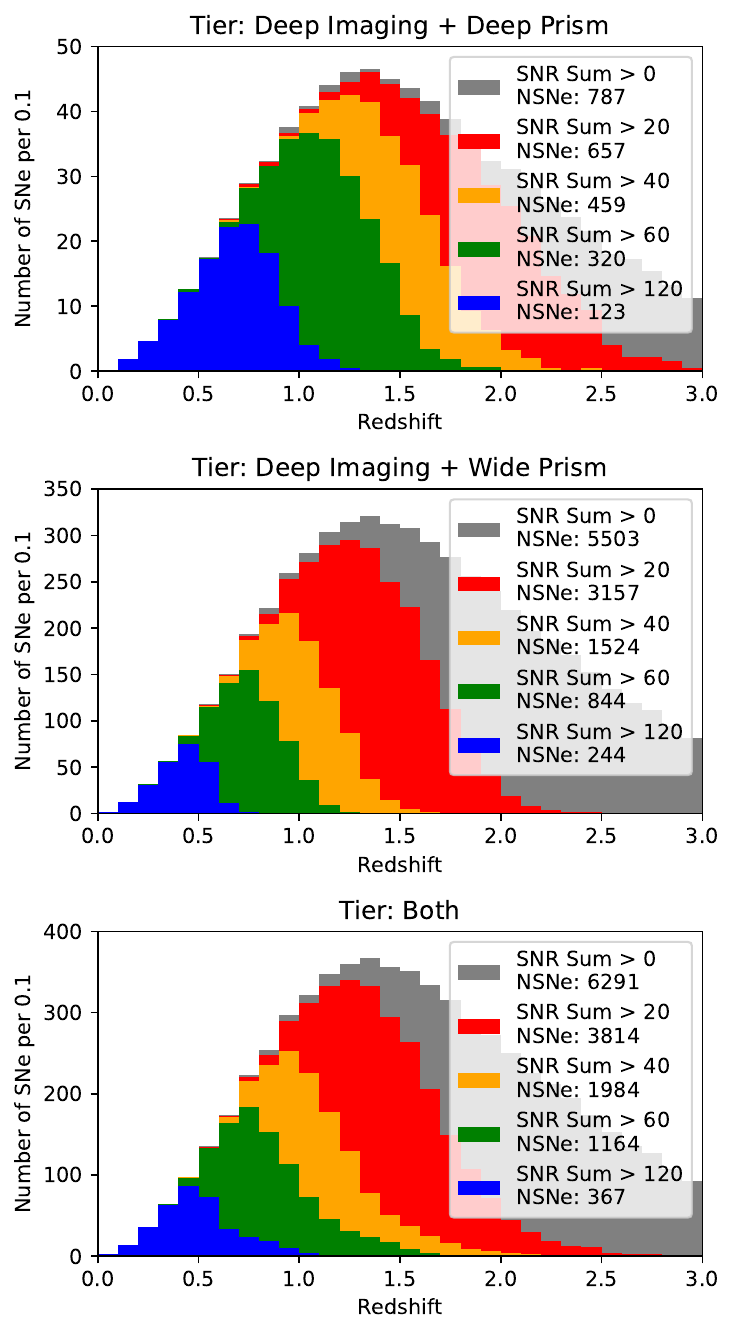}
    \caption{Quadrature sum of prism spectra S/N values. The colors show different S/N thresholds and we average over five surveys to reduce the sampling noise. Each {\bf panel} shows a different tier and the {\bf bottom panel} shows the combination of all tiers.}
    \label{fig:SpectraSNR}
\end{figure}

\begin{figure*}[htbp]
    \centering
    \includegraphics[width=0.8\textwidth]{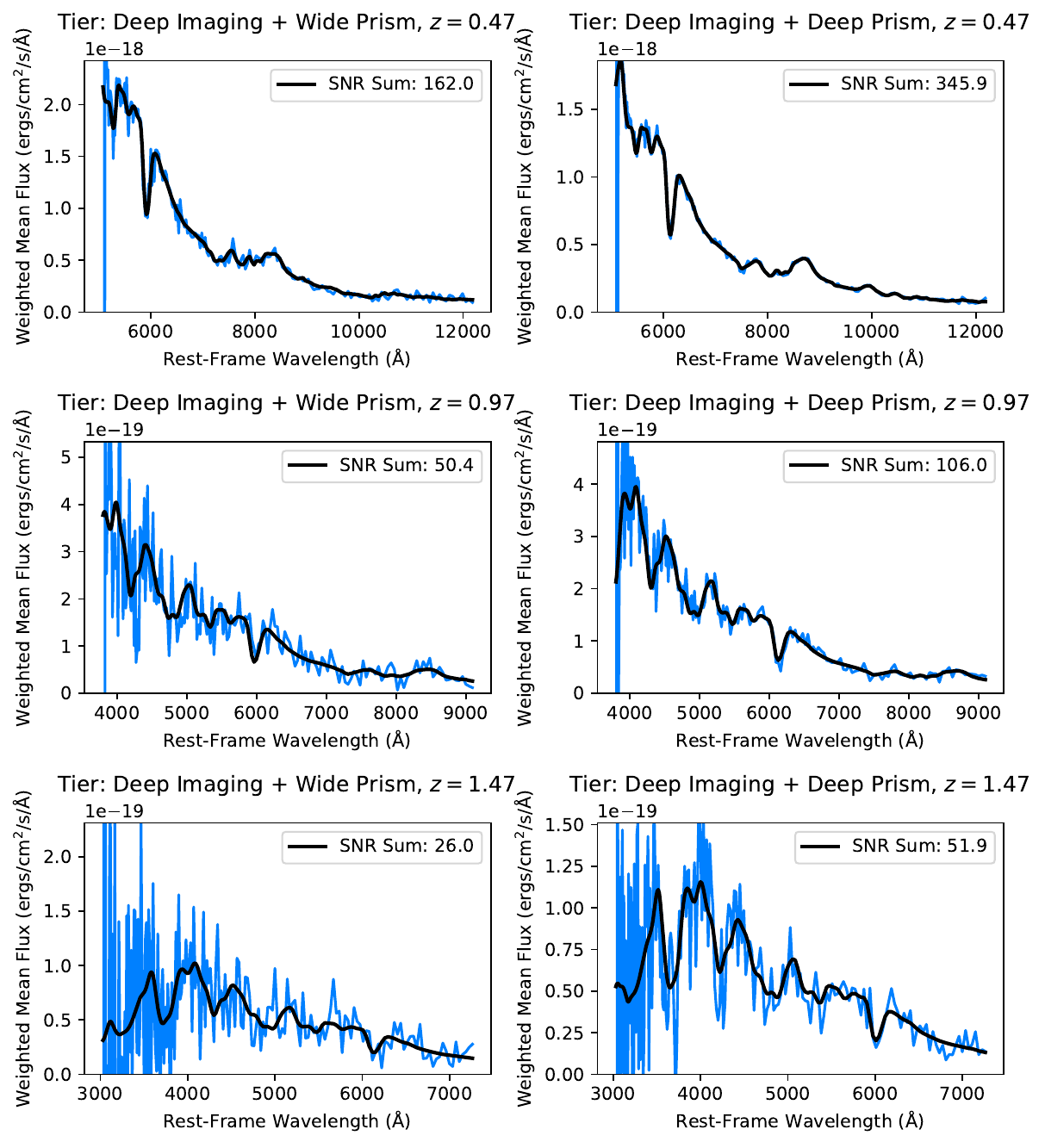}
    \caption{As in Figure~\ref{fig:MedianLCs}, median S/N prism spectra for both prism tiers and three different redshifts. For display purposes only, we combine the full prism time series into one spectrum (weighting by (S/N)$^2$) although our analysis treats each spectrum in the time series individually. This figure shows the combined spectrum with noise (blue lines) and without noise (thicker black lines). The total quadrature S/N (``SNR Sum'') of all wavelengths and spectra is given in the legend; these can be compared to the S/N values in Figures~\ref{fig:twins} and \ref{fig:SpectraSNR}.}
    \label{fig:SampleSpectra}
\end{figure*}

Figure~\ref{fig:ImagingSNR} shows the quadrature \secondhalfsentence The nearby SNe and Rubin SNe complement the redshift range of Roman SNe. Figure~\ref{fig:MedianLCs} shows light curves as a function of redshift for the Rubin DDF and both Roman tiers. \lightcurvescomments Similarly, Figure~\ref{fig:SpectraSNR} shows the quadrature combined S/N counts for the prism spectra and Figure~\ref{fig:SampleSpectra} shows the median prism observation as a function of redshift for each tier.

\begin{figure}
    \centering
    \includegraphics[width=0.5\textwidth]{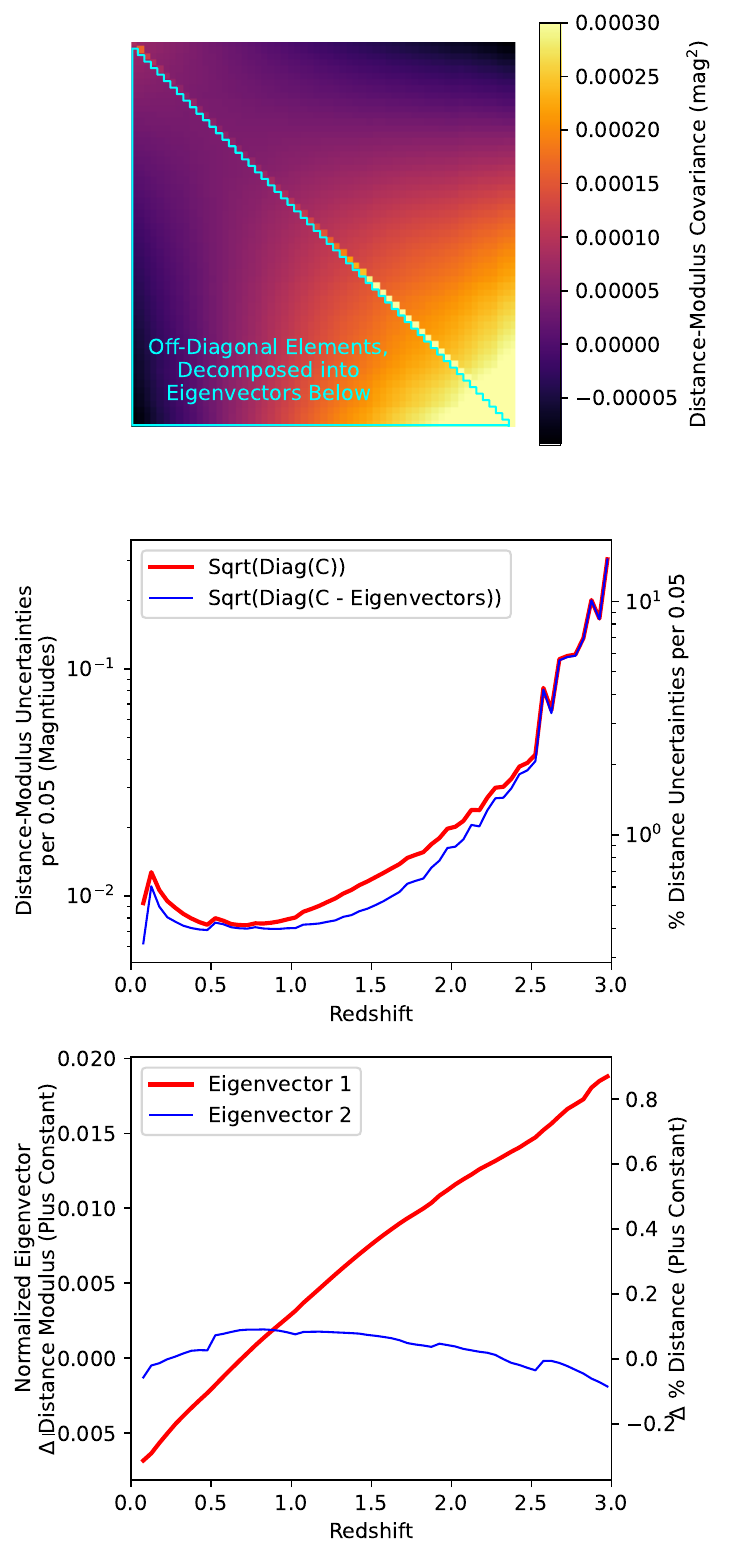}
    \caption{{\bf Top panel}: a visualization of the distance-modulus covariance matrix from the \CCSsurvey + Rubin DDF as a function of redshift in bins of 0.05. {\bf Middle panel}: distance uncertainties as a function of redshift. The left axis shows distance modulus; the right axis shows \% distance. The {\bf bottom panel} shows the eigenvector decomposition (up to an arbitrary additive constant degenerate with absolute magnitude) of the off-diagonal uncertainties again in distance modulus and \% distance. Most of the power is in the first eigenvector, which is roughly linear with redshift.}
    \label{fig:decomposition}
\end{figure}

Finally, we examine the distance-modulus covariance matrix for the \CCSsurvey (including Rubin DDF SNe and evaluated with the NIR-lower \scatter model). The top panel of Figure~\ref{fig:decomposition} shows the distance-modulus uncertainties (the square root of the values along the diagonal of the distance-modulus covariance matrix) as well as the \% distance uncertainties (right-hand axis). We decompose the off-diagonal part of the covariance matrix into the first two eigenvectors (bottom panel). These are normalized such that the outer product of the eigenvector with itself explains the matrix (up to an arbitrary additive constant degenerate with absolute magnitude). Most of the off-diagonal correlations are captured by the first eigenvector (thicker red line), which is surprisingly linear in redshift with an amplitude of about 1.2\% in distance between redshift 0 and 3. The second eigenvector (thinner blue line) has a much smaller impact, with some jumps due to changes in filters falling out of the rest-frame wavelength range modeled. After subtracting these eigenvectors, the remaining uncertainty is plotted in the top panel (thin blue line). The distance uncertainty per 0.05 in redshift is below 1\% up to redshift~$\sim 2$.

\section{Conclusions} \label{sec:Conclusion}

This work was part of a \RomanST SN Project Infrastructure Team effort that sought to help the Core Community Survey (CCS) Committee optimize the \RomanST High Latitude Time Domain Survey (HLTDS) for SN cosmology. Our primary results come from simulating 1,000 survey strategies for the  varying cadence, filter choice, fration of time in imaging and spectroscopic tiers, and targeted redshift to set the exposure times. We required our results to obey other constraints like a fixed total time (nominally 0.5 years) and minimum length of exposure time (to avoid being dominated by read noise and overheads).  We simulated each survey with and without a very conservative version of the Rubin Observatory Deep Drilling Field (DDF) SNe, for a total of 2,000 simulations. We also simulate different survey lengths to investigate the impact of varying the assumed 0.5 years.

For each simulation, we perform Fisher-matrix calculations to obtain the Dark Energy Task Force Figure of Merit (FoM, \citealt{Albrecht2006}) based on the $w_0$-$w_a$ constraints. We show how our Fisher-matrix FoM results are sensitive to the assumed calibration uncertainties and the SN \scatter model, and we run every survey with two different \scatter models (``NIR lower'' based on a color-scatter approach and ``Twins lower'' based on the twin-SN analysis). This is possibly the largest collection of FoM values in one analysis. We consider our FoM values reasonable for a relative ranking of the surveys, but there are unknowns at this point for interpreting the FoM values as absolute numbers.

In short, we find the following points.
\begin{itemize}
    \item Because of the CMB S4 delay, we evaluated FoM using a Planck shift-parameter constraint (0.26\% as opposed to the 0.2\% the \RomanST Science Definition Team assumed). The FoM drops $9\% \pm 0.8\%$ (survey-to-survey variation) with the 0.26\% shift-parameter constraint (Section~\ref{sec:DETF}).

    \item Our choice of \scatter model makes a large difference without considering Rubin DDF SNe; with NIR lower preferring a larger Wide-imaging tier and less prism time, and Twins lower the opposite. However, most of the difference between these models goes away when Rubin DDF SNe are also considered.  (Figure~\ref{fig:scatter}).

    \item The impact of the Rubin DDF SNe on FoM is large, although it may be somewhat overstated here as the nearby SNe are assumed to be calibrated to the Rubin Observatory (Figure~\ref{fig:scatter}).
    
    \item The HLTDS is not necessarily systematics-dominated at all redshifts; the FoM still increases for surveys at least up to 1 year of total time (Figure~\ref{fig:FoMTime}).
    \item Training the mean SN model matters at the factor of two level on FoM, so it is important to take this into account (which our Fisher-matrix calculations do). Photometric calibration uncertainties are the dominant systematic uncertainties that we model, and these have much more impact when the training of the mean SN model is also included (Figure~\ref{fig:FoMvsUnc}).

    \item Slower cadence increases FoM at fixed depth/day, especially for the Wide-tier which is proportionately more affected by read noise and overheads (Figure~\ref{fig:OtherParams}). Interlaced surveys, where roughly half the filters are observed in each visit, were rapidly found to be better than all-filters-every-visit because they effectively double the cadence. Thus, these are the only cadences considered in this work.

    \item The proposed \CCSsurvey (30\% time in Wide imaging targeting $z=0.9$, 50\% time in Deep imaging targeting $z=1.7$, 20\% time in prism, see Table~\ref{tab:CCSsurvey}), would produce $\sim \mathrm{15,000}$ SNe~Ia with high-quality light curves (Figures~\ref{fig:ImagingSNR} and \ref{fig:MedianLCs}) and time series ranging from subclassifiable ($\sim \mathrm{1,000}$~SNe) to lower S/N for redshift measurements ($\sim \mathrm{4,000}$~SNe), Figures~\ref{fig:SpectraSNR} and \ref{fig:SampleSpectra}). The \CCSsurvey is a reasonable choice based on our analysis.
\end{itemize}

A Fisher-matrix analysis cannot easily capture every term in a real cosmology analysis, and there are real uncertainties that will not be addressed until after \RomanST is on sky (SN rates, the size and form of calibration uncertainties, any unexpected ``unknown unknowns'' in SNe~Ia as a function of redshift). In any case, we believe we have come up with a robust ranking of surveys (especially when Rubin DDF SNe are included) and that the proposed \CCSsurvey will deliver a generation-defining cosmological measurement.

\begin{acknowledgments}
Support for this work was provided by NASA under contract 80NSSC24M0023 through the Roman Supernova Project Infrastructure Team. The material is based on work supported by NASA under award number 80GSFC24M0006. The technical support and advanced computing resources from University of Hawaii Information Technology Services – Research Cyberinfrastructure, funded in part by the National Science Foundation CC* awards \#2201428 and \#2232862 are gratefully acknowledged.
\end{acknowledgments}

\software{Matplotlib \citep{matplotlib}, 
Numpy \citep{numpy2020}, 
Python, 
SciPy \citep{scipy}, 
SNCosmo \citep{Barbary2016a}
}

\appendix

\section{Motivation for S/N Selection Cut} \label{sec:selectioncuts}

This Appendix motivates selecting on the total quadrature sum of S/N which we refer to as ``\SNRSUM'' (Equation~\ref{eq:SNRSum}). First, the motivation for defining \SNRSUM is that it is the S/N of a single amplitude parameter if the model (or template) is assumed to be correct, the data have little scatter about the model, and one has independent, Gaussian uncertainties. If one writes the $\chi^2$ as
\begin{align}
\chi^2 =& \sum_i \left[ \frac{\mathrm{data}_i - A\,\mathrm{model}_i}{\mathrm{uncertainty}_i} \right]^2 
 \nonumber \\
=& \sum_i \frac{\mathrm{data}_i^2}{\mathrm{uncertainty}_i^2}
\;-\; 2\,A \sum_i \frac{\mathrm{data}_i \,\mathrm{model}_i}{\mathrm{uncertainty}_i^2}
\;+\; A^2 \sum_i \frac{\mathrm{model}_i^2}{\mathrm{uncertainty}_i^2} \;, \nonumber
\end{align}
then one can take the derivative with respect to the model amplitude $A$ to find that the optimum $\hat{A}$ is
\begin{equation*}
    \hat{A} = \left[\sum_i \frac{\mathrm{data}_i \,\mathrm{model}_i}{\mathrm{uncertainty}_i^2} \right]/ \left[ \sum_i \frac{\mathrm{model}_i^2}{\mathrm{uncertainty}_i^2} \right] \;.
\end{equation*}
Taking one half the second derivative with respect to $A$ gives the inverse variance
\begin{equation*}
    \sigma_A^{-2} = \sum_i \frac{\mathrm{model}_i^2}{\mathrm{uncertainty}_i^2} \;.
\end{equation*}
Thus, the S/N of the amplitude parameter (best fit divided by $\sqrt{\mathrm{variance}}$) is 
\begin{equation*}
        \frac{\hat{A}}{\sigma_A} = \left[\sum_i \frac{\mathrm{data}_i \,\mathrm{model}_i}{\mathrm{uncertainty}_i^2} \right]/ \sqrt{ \sum_i \frac{\mathrm{model}_i^2}{\mathrm{uncertainty}_i^2}} \;.
\end{equation*}
Note that arbitrarily rescaling the model will cancel out of the S/N, as expected. If the data are not scattered by their uncertainties, and the model is the correct one, then data and model will be similar (up to an overall arbitrary scaling constant) and one can simply write the S/N of the amplitude as 
\begin{equation*}
    \sqrt{\sum_i \frac{\mathrm{data}_i^2}{\mathrm{uncertainty}_i^2}} \;,
\end{equation*}
i.e., Equation~\ref{eq:SNRSum}.

As noted in Section~\ref{sec:Fisher}, we require \SNRSUM~$> 40$ (summed in quadrature over all light-curve points) to include a SN our analysis and this is only an approximation of what the actual selection cuts will be. However, this is a good approximation for two reasons.

\begin{figure}
    \centering
    \includegraphics[width=0.5\linewidth]{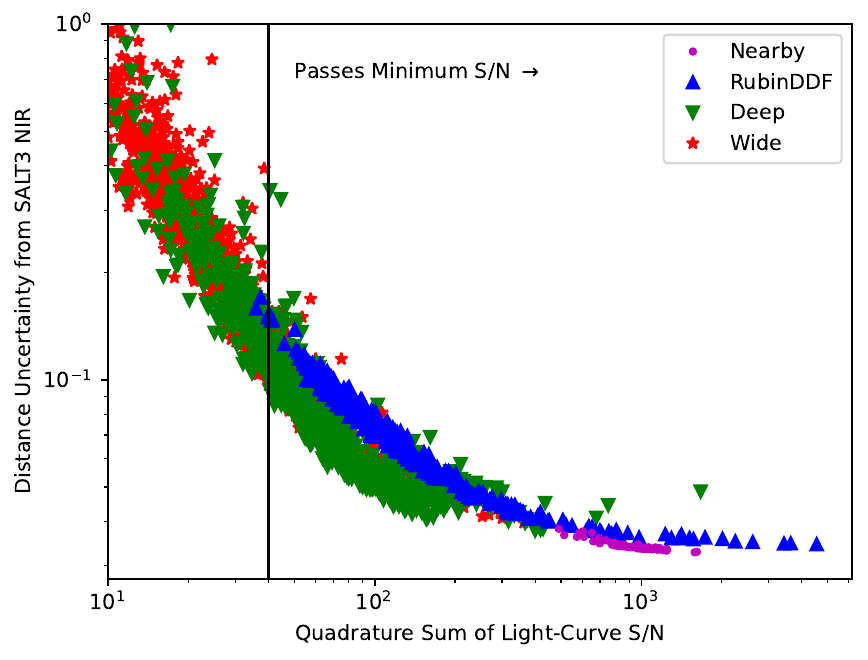}
    \caption{Distance-modulus uncertainty (assuming linear standardization, \citealt{Tripp1998}) vs. the total quadrature sum of light-curve S/N (\SNRSUM). The consistent depth and cadence produce a relation with modest scatter, even over orders of magnitude of S/N.}
    \label{fig:dmuvsSN}
\end{figure}

\begin{itemize}
    \item There are some biases that scale as (\SNRSUM)$^{-2}$. For example, photometric biases due to centroiding the SN, Malmquist bias in which SNe are selected, Eddington bias in the light-curve fitting and training, and non-Ia contamination. Some of these will even be influenced by the S/N in a single filter. Thus for the actual analysis, there will come a crossover point when accepting higher and higher redshift SNe where the statistical weight of each new SN added decreases as (\SNRSUM)$^2$ while the bias increases as (\SNRSUM)$^{-2}$.  It is unclear how well these biases can be corrected, and these terms are difficult to include in a Fisher-matrix analysis. These considerations motivate our use of a simple S/N cut in this work as an approximation to the true treatment.

    \item With a consistent cadence and depth (at least compared to prior surveys), \SNRSUM is closely related to the distance-modulus uncertainty, as shown for a random subset of SNe from the \CCSsurvey in Figure~\ref{fig:dmuvsSN} which we fit with SALT3-NIR.
\end{itemize}

Our chosen S/N cut of \SNRSUM~$> 40$ in quadrature across the whole light curve (or $\gtrsim 15$--$20$ per band) corresponds to biases $\mathcal{O}(40^{-2}$--$15^{-2}) = \mathcal{O}$(0.7--5~mmag) depending on whether one considers total S/N or per-filter S/N for the filters in which the SN is brightest. Thus, we are cutting at the point where the biases would matter for a real analysis, but they are not so large that correcting them to the level needed for \RomanST seems implausible.

\section{Exposure Times}

\startlongtable
\begin{deluxetable*}{ccccccccccccccc}
\tablecaption{Exposure time by filter, targeted redshift, and cadence.}
    \tablehead{\colhead{$z$} & 
    \colhead{2d} & 
    \colhead{3d} & 
    \colhead{4d} & 
    \colhead{5d} & 
    \colhead{6d} & 
    \colhead{7d} & 
    \colhead{8d} & 
    \colhead{9d} & 
    \colhead{10d} & 
    \colhead{11d} & 
    \colhead{12d} & 
    \colhead{13d} & 
    \colhead{14d} & 
    \colhead{15d}
    }
\startdata
\\ \multicolumn{15}{c}{\FRR} \\
$z=0.20$ & 60.0 & 60.0 & 60.0 & 60.0 & 60.0 & 60.0 & 60.0 & 60.0 & 60.0 & 60.0 & 60.0 & 60.0 & 60.0 & 60.0 \\ 
$z=0.30$ & 60.0 & 60.0 & 60.0 & 60.0 & 60.0 & 60.0 & 60.0 & 60.0 & 60.0 & 60.0 & 60.0 & 60.0 & 60.0 & 60.0 \\ 
$z=0.40$ & 60.0 & 60.0 & 60.0 & 60.0 & 60.0 & 60.0 & 60.0 & 60.0 & 60.0 & 60.0 & 60.0 & 60.0 & 60.0 & 60.0 \\ 
$z=0.50$ & 60.0 & 60.0 & 60.0 & 60.0 & 60.0 & 60.0 & 60.0 & 60.0 & 60.0 & 60.0 & 60.0 & 60.0 & 60.0 & 60.0 \\ 
$z=0.60$ & 60.0 & 60.0 & 60.0 & 60.0 & 60.0 & 60.0 & 60.0 & 60.0 & 60.0 & 60.0 & 60.0 & 60.0 & 60.0 & 60.0 \\ 
$z=0.70$ & 60.0 & 60.0 & 60.0 & 60.0 & 60.0 & 60.0 & 60.0 & 60.0 & 60.0 & 61.6 & 64.8 & 67.9 & 70.8 & 74.1 \\ 
$z=0.80$ & 60.0 & 60.0 & 60.0 & 60.0 & 62.0 & 67.6 & 73.0 & 78.1 & 83.2 & 88.2 & 93.4 & 98.3 & 103.0 & 108.0 \\ 
$z=0.90$ & 60.0 & 60.0 & 60.0 & 60.0 & 62.0 & 67.6 & 73.0 & 78.1 & 83.2 & 88.2 & 93.4 & 98.3 & 103.0 & 108.0 \\ 
$z=1.00$ & 60.0 & 60.0 & 60.0 & 60.0 & 62.0 & 67.6 & 73.0 & 78.1 & 83.2 & 88.2 & 93.4 & 98.3 & 103.0 & 108.0 \\ 
$z=1.10$ & 60.0 & 60.0 & 60.0 & 60.0 & 62.0 & 67.6 & 73.0 & 78.1 & 83.2 & 88.2 & 93.4 & 98.3 & 103.0 & 108.0 \\ 
$z=1.20$ & 60.0 & 60.0 & 60.0 & 60.0 & 62.0 & 67.6 & 73.0 & 78.1 & 83.2 & 88.2 & 93.4 & 98.3 & 103.0 & 108.0 \\ 
$z=1.30$ & 60.0 & 60.0 & 60.0 & 60.0 & 62.0 & 67.6 & 73.0 & 78.1 & 83.2 & 88.2 & 93.4 & 98.3 & 103.0 & 108.0 \\ 
$z=1.40$ & 60.0 & 60.0 & 60.0 & 60.0 & 62.0 & 67.6 & 73.0 & 78.1 & 83.2 & 88.2 & 93.4 & 98.3 & 103.0 & 108.0 \\ 
$z=1.50$ & 60.0 & 60.0 & 60.0 & 60.0 & 62.0 & 67.6 & 73.0 & 78.1 & 83.2 & 88.2 & 93.4 & 98.3 & 103.0 & 108.0 \\ 
$z=1.60$ & 60.0 & 60.0 & 60.0 & 60.0 & 62.0 & 67.6 & 73.0 & 78.1 & 83.2 & 88.2 & 93.4 & 98.3 & 103.0 & 108.0 \\ 
$z=1.70$ & 60.0 & 60.0 & 60.0 & 60.0 & 62.0 & 67.6 & 73.0 & 78.1 & 83.2 & 88.2 & 93.4 & 98.3 & 103.0 & 108.0 \\ 
$z=1.80$ & 60.0 & 60.0 & 60.0 & 60.0 & 62.0 & 67.6 & 73.0 & 78.1 & 83.2 & 88.2 & 93.4 & 98.3 & 103.0 & 108.0 \\ 
$z=1.90$ & 60.0 & 60.0 & 60.0 & 60.0 & 62.0 & 67.6 & 73.0 & 78.1 & 83.2 & 88.2 & 93.4 & 98.3 & 103.0 & 108.0 \\ 
$z=2.00$ & 60.0 & 60.0 & 60.0 & 60.0 & 62.0 & 67.6 & 73.0 & 78.1 & 83.2 & 88.2 & 93.4 & 98.3 & 103.0 & 108.0 \\ 
$z=2.10$ & 60.0 & 60.0 & 60.0 & 60.0 & 62.0 & 67.6 & 73.0 & 78.1 & 83.2 & 88.2 & 93.4 & 98.3 & 103.0 & 108.0 \\ 
$z=2.20$ & 60.0 & 60.0 & 60.0 & 60.0 & 62.0 & 67.6 & 73.0 & 78.1 & 83.2 & 88.2 & 93.4 & 98.3 & 103.0 & 108.0 \\ 
$z=2.30$ & 60.0 & 60.0 & 60.0 & 60.0 & 62.0 & 67.6 & 73.0 & 78.1 & 83.2 & 88.2 & 93.4 & 98.3 & 103.0 & 108.0 \\ 
$z=2.40$ & 60.0 & 60.0 & 60.0 & 60.0 & 62.0 & 67.6 & 73.0 & 78.1 & 83.2 & 88.2 & 93.4 & 98.3 & 103.0 & 108.0 \\ 
$z=2.50$ & 60.0 & 60.0 & 60.0 & 60.0 & 62.0 & 67.6 & 73.0 & 78.1 & 83.2 & 88.2 & 93.4 & 98.3 & 103.0 & 108.0 \\ 
\hline
\multicolumn{15}{c}{\FZ} \\
$z=0.20$ & 60.0 & 60.0 & 60.0 & 60.0 & 60.0 & 60.0 & 60.0 & 60.0 & 60.0 & 60.0 & 60.0 & 60.0 & 60.0 & 60.0 \\ 
$z=0.30$ & 60.0 & 60.0 & 60.0 & 60.0 & 60.0 & 60.0 & 60.0 & 60.0 & 60.0 & 60.0 & 60.0 & 60.0 & 60.0 & 60.0 \\ 
$z=0.40$ & 60.0 & 60.0 & 60.0 & 60.0 & 60.0 & 60.0 & 60.0 & 60.0 & 60.0 & 60.0 & 60.0 & 60.0 & 60.0 & 60.0 \\ 
$z=0.50$ & 60.0 & 60.0 & 60.0 & 60.0 & 60.0 & 60.0 & 60.0 & 60.0 & 60.0 & 60.0 & 60.0 & 60.0 & 60.0 & 60.0 \\ 
$z=0.60$ & 60.0 & 60.0 & 60.0 & 60.0 & 60.0 & 60.0 & 60.0 & 60.0 & 60.0 & 60.0 & 60.0 & 60.0 & 60.0 & 61.8 \\ 
$z=0.70$ & 60.0 & 60.0 & 60.0 & 60.0 & 60.0 & 60.0 & 60.0 & 60.0 & 60.6 & 64.0 & 67.4 & 70.7 & 73.9 & 77.1 \\ 
$z=0.80$ & 60.0 & 60.0 & 60.0 & 60.0 & 60.0 & 62.6 & 67.5 & 72.2 & 76.8 & 81.4 & 85.9 & 90.2 & 94.9 & 99.0 \\ 
$z=0.90$ & 60.0 & 60.0 & 60.0 & 60.0 & 65.3 & 71.0 & 76.6 & 82.3 & 87.6 & 93.1 & 98.2 & 103.5 & 108.6 & 113.7 \\ 
$z=1.00$ & 60.0 & 60.0 & 60.3 & 67.8 & 74.9 & 81.6 & 88.3 & 94.9 & 101.2 & 107.8 & 114.0 & 120.4 & 126.6 & 132.9 \\ 
$z=1.10$ & 60.0 & 62.0 & 71.9 & 81.2 & 90.1 & 98.7 & 107.2 & 115.6 & 123.8 & 132.2 & 140.2 & 148.5 & 156.5 & 165.0 \\ 
$z=1.20$ & 60.5 & 74.1 & 86.4 & 98.0 & 109.3 & 120.3 & 131.2 & 141.8 & 152.6 & 163.5 & 174.0 & 184.6 & 195.4 & 206.1 \\ 
$z=1.30$ & 72.7 & 89.8 & 105.6 & 120.7 & 135.4 & 149.9 & 164.3 & 178.4 & 193.0 & 207.2 & 221.3 & 235.8 & 250.0 & 264.6 \\ 
$z=1.40$ & 94.0 & 118.0 & 140.4 & 162.4 & 184.1 & 205.7 & 227.0 & 248.6 & 270.4 & 291.9 & 313.4 & 335.4 & 356.7 & 378.9 \\ 
$z=1.50$ & 133.0 & 171.2 & 208.2 & 244.6 & 280.8 & 317.0 & 353.3 & 389.7 & 426.2 & 462.7 & 498.7 & 535.6 & 571.8 & 608.7 \\ 
$z=1.60$ & 133.0 & 171.2 & 208.2 & 244.6 & 280.8 & 317.0 & 353.3 & 389.7 & 426.2 & 462.7 & 498.7 & 535.6 & 571.8 & 608.7 \\ 
$z=1.70$ & 133.0 & 171.2 & 208.2 & 244.6 & 280.8 & 317.0 & 353.3 & 389.7 & 426.2 & 462.7 & 498.7 & 535.6 & 571.8 & 608.7 \\ 
$z=1.80$ & 133.0 & 171.2 & 208.2 & 244.6 & 280.8 & 317.0 & 353.3 & 389.7 & 426.2 & 462.7 & 498.7 & 535.6 & 571.8 & 608.7 \\ 
$z=1.90$ & 133.0 & 171.2 & 208.2 & 244.6 & 280.8 & 317.0 & 353.3 & 389.7 & 426.2 & 462.7 & 498.7 & 535.6 & 571.8 & 608.7 \\ 
$z=2.00$ & 133.0 & 171.2 & 208.2 & 244.6 & 280.8 & 317.0 & 353.3 & 389.7 & 426.2 & 462.7 & 498.7 & 535.6 & 571.8 & 608.7 \\ 
$z=2.10$ & 133.0 & 171.2 & 208.2 & 244.6 & 280.8 & 317.0 & 353.3 & 389.7 & 426.2 & 462.7 & 498.7 & 535.6 & 571.8 & 608.7 \\ 
$z=2.20$ & 133.0 & 171.2 & 208.2 & 244.6 & 280.8 & 317.0 & 353.3 & 389.7 & 426.2 & 462.7 & 498.7 & 535.6 & 571.8 & 608.7 \\ 
$z=2.30$ & 133.0 & 171.2 & 208.2 & 244.6 & 280.8 & 317.0 & 353.3 & 389.7 & 426.2 & 462.7 & 498.7 & 535.6 & 571.8 & 608.7 \\ 
$z=2.40$ & 133.0 & 171.2 & 208.2 & 244.6 & 280.8 & 317.0 & 353.3 & 389.7 & 426.2 & 462.7 & 498.7 & 535.6 & 571.8 & 608.7 \\ 
$z=2.50$ & 133.0 & 171.2 & 208.2 & 244.6 & 280.8 & 317.0 & 353.3 & 389.7 & 426.2 & 462.7 & 498.7 & 535.6 & 571.8 & 608.7 \\ 
\hline
\multicolumn{15}{c}{\FY} \\
$z=0.20$ & 60.0 & 60.0 & 60.0 & 60.0 & 60.0 & 60.0 & 60.0 & 60.0 & 60.0 & 60.0 & 60.0 & 60.0 & 60.0 & 60.0 \\ 
$z=0.30$ & 60.0 & 60.0 & 60.0 & 60.0 & 60.0 & 60.0 & 60.0 & 60.0 & 60.0 & 60.0 & 60.0 & 60.0 & 60.0 & 60.0 \\ 
$z=0.40$ & 60.0 & 60.0 & 60.0 & 60.0 & 60.0 & 60.0 & 60.0 & 60.0 & 60.0 & 60.0 & 60.0 & 60.0 & 60.0 & 60.0 \\ 
$z=0.50$ & 60.0 & 60.0 & 60.0 & 60.0 & 60.0 & 60.0 & 60.0 & 60.0 & 60.0 & 60.0 & 60.0 & 60.1 & 62.7 & 65.4 \\ 
$z=0.60$ & 60.0 & 60.0 & 60.0 & 60.0 & 60.0 & 60.0 & 60.0 & 60.0 & 61.4 & 64.9 & 68.4 & 71.8 & 75.0 & 78.3 \\ 
$z=0.70$ & 60.0 & 60.0 & 60.0 & 60.0 & 60.0 & 60.0 & 60.2 & 64.3 & 68.2 & 72.2 & 76.1 & 80.1 & 84.0 & 87.6 \\ 
$z=0.80$ & 60.0 & 60.0 & 60.0 & 60.0 & 60.5 & 65.9 & 71.0 & 76.1 & 81.2 & 86.2 & 91.0 & 95.9 & 100.5 & 105.6 \\ 
$z=0.90$ & 60.0 & 60.0 & 60.0 & 64.9 & 71.8 & 78.4 & 84.8 & 91.1 & 97.4 & 103.6 & 109.7 & 116.0 & 122.1 & 128.1 \\ 
$z=1.00$ & 60.0 & 60.0 & 66.7 & 75.3 & 83.5 & 91.4 & 99.2 & 106.9 & 114.6 & 122.1 & 129.8 & 137.3 & 144.8 & 152.4 \\ 
$z=1.10$ & 60.0 & 64.8 & 75.4 & 85.4 & 95.0 & 104.4 & 113.6 & 122.8 & 132.0 & 140.8 & 150.0 & 158.9 & 168.0 & 177.0 \\ 
$z=1.20$ & 61.0 & 74.9 & 87.6 & 99.6 & 111.4 & 123.1 & 134.4 & 145.8 & 157.0 & 168.5 & 179.8 & 191.1 & 202.4 & 213.9 \\ 
$z=1.30$ & 64.7 & 79.7 & 93.6 & 106.7 & 119.5 & 132.2 & 144.5 & 156.8 & 169.2 & 181.5 & 193.7 & 205.9 & 218.4 & 230.7 \\ 
$z=1.40$ & 70.4 & 87.2 & 102.6 & 117.6 & 132.1 & 146.4 & 160.6 & 175.0 & 189.2 & 203.5 & 217.7 & 231.9 & 246.1 & 260.7 \\ 
$z=1.50$ & 79.4 & 99.1 & 117.5 & 135.3 & 152.8 & 170.1 & 187.4 & 204.5 & 221.6 & 239.1 & 256.1 & 273.5 & 290.9 & 308.1 \\ 
$z=1.60$ & 90.1 & 113.2 & 135.1 & 156.4 & 177.4 & 198.4 & 219.4 & 240.5 & 261.4 & 282.3 & 303.6 & 324.7 & 345.8 & 367.2 \\ 
$z=1.70$ & 105.2 & 133.7 & 161.0 & 188.1 & 214.9 & 241.6 & 268.5 & 295.2 & 322.0 & 348.9 & 376.1 & 403.0 & 430.4 & 457.5 \\ 
$z=1.80$ & 122.0 & 156.7 & 190.6 & 223.9 & 257.5 & 291.1 & 324.6 & 358.0 & 392.0 & 425.9 & 459.8 & 493.7 & 528.1 & 561.6 \\ 
$z=1.90$ & 154.9 & 202.9 & 249.9 & 297.0 & 343.8 & 390.6 & 437.6 & 484.9 & 532.2 & 579.3 & 626.2 & 673.9 & 721.0 & 768.6 \\ 
$z=2.00$ & 206.1 & 277.3 & 348.6 & 420.2 & 491.6 & 563.4 & 635.7 & 707.8 & 780.6 & 853.4 & 925.9 & 997.9 & 1071.6 & 1143.9 \\ 
$z=2.10$ & 206.1 & 277.3 & 348.6 & 420.2 & 491.6 & 563.4 & 635.7 & 707.8 & 780.6 & 853.4 & 925.9 & 997.9 & 1071.6 & 1143.9 \\ 
$z=2.20$ & 206.1 & 277.3 & 348.6 & 420.2 & 491.6 & 563.4 & 635.7 & 707.8 & 780.6 & 853.4 & 925.9 & 997.9 & 1071.6 & 1143.9 \\ 
$z=2.30$ & 206.1 & 277.3 & 348.6 & 420.2 & 491.6 & 563.4 & 635.7 & 707.8 & 780.6 & 853.4 & 925.9 & 997.9 & 1071.6 & 1143.9 \\ 
$z=2.40$ & 206.1 & 277.3 & 348.6 & 420.2 & 491.6 & 563.4 & 635.7 & 707.8 & 780.6 & 853.4 & 925.9 & 997.9 & 1071.6 & 1143.9 \\ 
$z=2.50$ & 206.1 & 277.3 & 348.6 & 420.2 & 491.6 & 563.4 & 635.7 & 707.8 & 780.6 & 853.4 & 925.9 & 997.9 & 1071.6 & 1143.9 \\ 
\hline
\multicolumn{15}{c}{\FJ} \\
$z=0.20$ & 60.0 & 60.0 & 60.0 & 60.0 & 60.0 & 60.0 & 60.0 & 60.0 & 60.0 & 60.0 & 60.0 & 60.0 & 60.0 & 60.0 \\ 
$z=0.30$ & 60.0 & 60.0 & 60.0 & 60.0 & 60.0 & 60.0 & 60.0 & 60.0 & 60.0 & 60.0 & 60.0 & 60.0 & 60.0 & 60.0 \\ 
$z=0.40$ & 60.0 & 60.0 & 60.0 & 60.0 & 60.0 & 60.0 & 60.0 & 60.0 & 60.0 & 60.0 & 63.1 & 66.3 & 69.2 & 72.3 \\ 
$z=0.50$ & 60.0 & 60.0 & 60.0 & 60.0 & 60.0 & 60.1 & 64.6 & 69.1 & 73.6 & 77.9 & 82.3 & 86.6 & 91.0 & 95.1 \\ 
$z=0.60$ & 60.0 & 60.0 & 60.0 & 63.6 & 70.3 & 76.9 & 83.2 & 89.3 & 95.6 & 101.4 & 107.5 & 113.6 & 119.6 & 125.4 \\ 
$z=0.70$ & 60.0 & 60.0 & 68.6 & 77.6 & 86.3 & 94.8 & 103.0 & 111.2 & 119.6 & 127.6 & 135.8 & 143.8 & 152.0 & 159.9 \\ 
$z=0.80$ & 60.0 & 62.8 & 73.1 & 82.9 & 92.3 & 101.5 & 110.6 & 119.5 & 128.4 & 137.3 & 146.2 & 155.0 & 163.8 & 172.5 \\ 
$z=0.90$ & 60.0 & 69.5 & 81.2 & 92.4 & 103.2 & 113.8 & 124.3 & 134.8 & 145.2 & 155.5 & 165.8 & 176.3 & 186.8 & 196.8 \\ 
$z=1.00$ & 60.0 & 70.7 & 82.8 & 94.3 & 105.4 & 116.2 & 126.9 & 137.5 & 148.2 & 158.6 & 169.4 & 179.9 & 190.4 & 201.0 \\ 
$z=1.10$ & 61.6 & 75.9 & 89.0 & 101.5 & 113.6 & 125.6 & 137.6 & 149.4 & 161.2 & 173.1 & 184.8 & 196.6 & 208.6 & 220.5 \\ 
$z=1.20$ & 66.5 & 82.3 & 97.0 & 111.0 & 124.7 & 138.2 & 151.7 & 165.1 & 178.4 & 191.8 & 205.2 & 218.7 & 232.1 & 245.4 \\ 
$z=1.30$ & 75.3 & 93.8 & 111.0 & 127.7 & 144.1 & 160.3 & 176.3 & 192.6 & 208.8 & 224.6 & 241.0 & 257.1 & 273.3 & 289.5 \\ 
$z=1.40$ & 82.9 & 104.1 & 124.0 & 143.5 & 162.8 & 182.0 & 201.0 & 220.3 & 239.6 & 258.7 & 277.9 & 297.4 & 316.4 & 336.3 \\ 
$z=1.50$ & 89.9 & 113.6 & 136.2 & 158.1 & 180.0 & 201.9 & 223.7 & 245.7 & 267.8 & 289.7 & 311.8 & 334.1 & 356.2 & 378.6 \\ 
$z=1.60$ & 100.8 & 128.3 & 154.6 & 180.4 & 206.0 & 231.6 & 257.1 & 282.6 & 308.4 & 334.2 & 359.5 & 385.6 & 411.3 & 437.1 \\ 
$z=1.70$ & 108.7 & 139.1 & 168.5 & 197.4 & 226.2 & 255.2 & 284.0 & 313.2 & 342.4 & 371.8 & 400.6 & 430.3 & 459.8 & 489.0 \\ 
$z=1.80$ & 114.0 & 146.5 & 177.5 & 208.6 & 239.3 & 269.9 & 300.6 & 331.6 & 362.2 & 392.9 & 424.1 & 455.0 & 486.1 & 517.2 \\ 
$z=1.90$ & 119.0 & 153.5 & 187.1 & 220.4 & 253.8 & 287.4 & 321.1 & 354.8 & 389.0 & 423.3 & 457.4 & 491.7 & 526.1 & 560.1 \\ 
$z=2.00$ & 133.7 & 174.2 & 214.2 & 253.8 & 293.8 & 333.8 & 373.8 & 414.0 & 454.2 & 494.6 & 535.2 & 575.6 & 616.3 & 657.0 \\ 
$z=2.10$ & 143.2 & 186.8 & 229.5 & 272.3 & 315.0 & 358.0 & 401.1 & 444.4 & 488.2 & 532.0 & 575.5 & 619.1 & 663.3 & 706.5 \\ 
$z=2.20$ & 165.7 & 219.4 & 272.7 & 326.1 & 379.7 & 433.3 & 487.2 & 541.3 & 595.4 & 649.7 & 703.7 & 758.4 & 812.8 & 867.3 \\ 
$z=2.30$ & 196.2 & 264.5 & 332.6 & 400.5 & 469.4 & 538.2 & 607.2 & 676.3 & 745.6 & 814.7 & 883.9 & 953.9 & 1023.1 & 1092.6 \\ 
$z=2.40$ & 223.8 & 305.9 & 388.2 & 471.4 & 554.5 & 638.3 & 722.4 & 806.4 & 890.4 & 975.5 & 1060.3 & 1144.5 & 1229.2 & 1314.3 \\ 
$z=2.50$ & 281.5 & 390.7 & 500.2 & 610.8 & 721.3 & 832.6 & 943.7 & 1055.0 & 1166.2 & 1277.3 & 1388.9 & 1501.0 & 1612.0 & 1724.1 \\ 
\hline
\multicolumn{15}{c}{\FH} \\
$z=0.20$ & 60.0 & 60.0 & 60.0 & 60.0 & 60.0 & 60.0 & 60.0 & 60.0 & 60.0 & 60.0 & 60.0 & 60.0 & 60.0 & 60.0 \\ 
$z=0.30$ & 60.0 & 60.0 & 60.0 & 60.0 & 60.0 & 60.0 & 60.6 & 64.8 & 69.0 & 73.0 & 77.0 & 81.1 & 85.1 & 88.8 \\ 
$z=0.40$ & 60.0 & 60.0 & 62.3 & 70.4 & 78.1 & 85.5 & 93.0 & 100.1 & 107.2 & 114.4 & 121.4 & 128.4 & 135.5 & 142.5 \\ 
$z=0.50$ & 60.0 & 70.0 & 82.0 & 93.3 & 104.4 & 115.2 & 125.8 & 136.6 & 147.2 & 157.7 & 168.5 & 178.9 & 189.6 & 200.1 \\ 
$z=0.60$ & 65.4 & 80.9 & 95.2 & 109.0 & 122.4 & 135.7 & 149.0 & 162.0 & 175.0 & 188.3 & 201.4 & 214.5 & 227.6 & 240.9 \\ 
$z=0.70$ & 72.7 & 90.7 & 107.5 & 123.7 & 139.6 & 155.4 & 171.2 & 186.8 & 202.6 & 218.2 & 234.0 & 249.6 & 265.7 & 281.4 \\ 
$z=0.80$ & 79.5 & 99.7 & 118.7 & 137.1 & 155.3 & 173.5 & 191.5 & 209.5 & 227.6 & 245.7 & 264.0 & 281.8 & 300.4 & 318.6 \\ 
$z=0.90$ & 93.0 & 117.8 & 141.4 & 164.9 & 188.2 & 211.3 & 234.4 & 257.6 & 280.6 & 304.0 & 327.6 & 350.7 & 374.4 & 397.5 \\ 
$z=1.00$ & 104.4 & 133.6 & 161.6 & 189.4 & 217.1 & 244.7 & 272.6 & 300.6 & 328.4 & 356.2 & 384.7 & 412.9 & 441.0 & 469.5 \\ 
$z=1.10$ & 118.3 & 152.8 & 186.6 & 219.9 & 253.3 & 287.0 & 320.6 & 354.2 & 388.2 & 422.4 & 456.2 & 490.4 & 524.7 & 558.3 \\ 
$z=1.20$ & 118.3 & 152.8 & 186.6 & 219.9 & 253.3 & 287.0 & 320.6 & 354.2 & 388.6 & 422.8 & 457.0 & 491.1 & 525.6 & 559.8 \\ 
$z=1.30$ & 127.5 & 165.8 & 203.2 & 240.4 & 277.4 & 314.6 & 351.8 & 389.3 & 427.0 & 464.6 & 501.8 & 539.8 & 577.4 & 615.3 \\ 
$z=1.40$ & 127.5 & 165.8 & 203.2 & 240.4 & 277.4 & 314.6 & 351.8 & 389.3 & 427.0 & 464.6 & 501.8 & 539.8 & 577.4 & 615.3 \\ 
$z=1.50$ & 127.5 & 165.8 & 203.2 & 240.4 & 277.4 & 314.6 & 351.8 & 389.3 & 427.0 & 464.6 & 501.8 & 539.8 & 577.4 & 615.3 \\ 
$z=1.60$ & 130.6 & 170.5 & 209.7 & 248.5 & 287.6 & 326.9 & 366.2 & 405.5 & 445.0 & 484.9 & 524.6 & 563.9 & 604.2 & 644.1 \\ 
$z=1.70$ & 136.6 & 178.9 & 220.6 & 262.4 & 304.1 & 345.9 & 388.0 & 430.4 & 472.8 & 515.0 & 557.3 & 600.3 & 642.6 & 685.5 \\ 
$z=1.80$ & 149.9 & 198.0 & 245.9 & 293.9 & 342.0 & 390.3 & 438.7 & 487.6 & 536.4 & 585.4 & 633.8 & 683.5 & 732.5 & 781.5 \\ 
$z=1.90$ & 169.4 & 226.6 & 283.6 & 341.1 & 398.4 & 456.4 & 514.4 & 572.4 & 630.4 & 689.0 & 747.6 & 805.7 & 864.6 & 923.1 \\ 
$z=2.00$ & 183.8 & 248.0 & 312.2 & 377.1 & 441.7 & 506.8 & 572.2 & 637.4 & 702.8 & 768.9 & 834.7 & 899.9 & 966.3 & 1032.0 \\ 
$z=2.10$ & 183.8 & 248.0 & 312.2 & 377.1 & 441.7 & 506.8 & 572.2 & 637.4 & 702.8 & 768.9 & 834.7 & 899.9 & 966.3 & 1032.0 \\ 
$z=2.20$ & 209.2 & 284.2 & 359.5 & 435.3 & 511.3 & 587.7 & 664.3 & 740.9 & 817.4 & 893.9 & 971.3 & 1048.3 & 1124.5 & 1202.4 \\ 
$z=2.30$ & 230.6 & 317.2 & 404.6 & 492.4 & 580.8 & 669.3 & 758.1 & 847.1 & 936.0 & 1025.0 & 1113.6 & 1203.5 & 1292.5 & 1381.8 \\ 
$z=2.40$ & 237.0 & 326.8 & 417.0 & 507.5 & 598.9 & 690.1 & 781.4 & 873.2 & 965.2 & 1057.3 & 1148.9 & 1240.7 & 1333.4 & 1424.7 \\ 
$z=2.50$ & 252.9 & 350.9 & 449.3 & 548.6 & 648.1 & 748.0 & 848.0 & 948.1 & 1048.2 & 1148.2 & 1248.2 & 1349.1 & 1449.3 & 1549.8 \\
\hline
\multicolumn{15}{c}{\FF} \\
$z=0.20$ & 60.0 & 60.0 & 60.0 & 60.0 & 63.2 & 68.9 & 74.2 & 79.6 & 84.8 & 90.0 & 95.3 & 100.1 & 105.3 & 110.4 \\ 
$z=0.30$ & 61.4 & 75.5 & 88.4 & 100.7 & 112.7 & 124.5 & 136.2 & 147.8 & 159.2 & 170.9 & 182.4 & 194.0 & 205.5 & 216.9 \\ 
$z=0.40$ & 91.6 & 115.4 & 138.1 & 160.2 & 182.3 & 204.1 & 225.9 & 247.9 & 270.0 & 292.2 & 313.9 & 336.4 & 358.4 & 381.0 \\ 
$z=0.50$ & 113.6 & 145.6 & 176.5 & 207.3 & 238.0 & 268.7 & 299.5 & 330.3 & 361.2 & 392.3 & 423.6 & 454.5 & 485.8 & 517.2 \\ 
$z=0.60$ & 141.5 & 185.2 & 228.0 & 271.0 & 313.8 & 357.0 & 400.3 & 443.9 & 487.8 & 531.7 & 575.5 & 619.3 & 663.6 & 707.1 \\ 
$z=0.70$ & 195.1 & 263.2 & 331.2 & 399.4 & 468.5 & 537.6 & 607.0 & 676.6 & 746.2 & 816.0 & 885.4 & 956.0 & 1025.6 & 1095.9 \\ 
$z=0.80$ & 238.1 & 326.5 & 415.4 & 504.6 & 595.0 & 685.0 & 775.4 & 866.0 & 957.0 & 1047.9 & 1138.3 & 1229.5 & 1321.0 & 1410.9 \\ 
$z=0.90$ & 248.3 & 341.5 & 435.1 & 529.5 & 623.9 & 718.9 & 814.2 & 909.5 & 1004.8 & 1100.7 & 1196.9 & 1292.5 & 1388.0 & 1484.4 \\ 
$z=1.00$ & 248.3 & 341.5 & 435.1 & 529.5 & 623.9 & 718.9 & 814.2 & 909.5 & 1004.8 & 1100.7 & 1196.9 & 1292.5 & 1388.0 & 1484.4 \\ 
$z=1.10$ & 248.3 & 341.5 & 435.1 & 529.5 & 623.9 & 718.9 & 814.2 & 909.5 & 1004.8 & 1100.7 & 1196.9 & 1292.5 & 1388.0 & 1484.4 \\ 
$z=1.20$ & 267.8 & 370.4 & 474.0 & 578.0 & 682.7 & 787.2 & 892.2 & 997.4 & 1103.2 & 1209.3 & 1315.0 & 1419.9 & 1526.6 & 1632.0 \\ 
$z=1.30$ & 337.2 & 474.7 & 613.0 & 752.4 & 891.4 & 1032.1 & 1172.5 & 1312.9 & 1453.2 & 1593.0 & 1734.5 & 1875.6 & 2015.4 & 2157.3 \\ 
$z=1.40$ & 417.2 & 594.7 & 773.3 & 952.9 & 1132.0 & 1312.8 & 1493.1 & 1673.5 & 1854.0 & 2033.9 & 2214.0 & 2395.9 & 2576.0 & 2756.4 \\ 
$z=1.50$ & 435.4 & 621.7 & 809.1 & 996.8 & 1186.1 & 1374.5 & 1563.0 & 1752.1 & 1941.6 & 2131.1 & 2320.3 & 2507.7 & 2698.9 & 2887.8 \\ 
$z=1.60$ & 435.4 & 621.7 & 809.1 & 996.8 & 1186.1 & 1374.5 & 1563.0 & 1752.1 & 1941.6 & 2131.1 & 2320.3 & 2507.7 & 2698.9 & 2887.8 \\ 
$z=1.70$ & 435.4 & 621.7 & 809.1 & 996.8 & 1186.1 & 1374.5 & 1563.0 & 1752.1 & 1941.6 & 2131.1 & 2320.3 & 2507.7 & 2698.9 & 2887.8 \\ 
$z=1.80$ & 435.4 & 621.7 & 809.1 & 996.8 & 1186.1 & 1374.5 & 1563.0 & 1752.1 & 1941.6 & 2131.1 & 2320.3 & 2507.7 & 2698.9 & 2887.8 \\ 
$z=1.90$ & 435.4 & 621.7 & 809.1 & 996.8 & 1186.1 & 1374.5 & 1563.0 & 1752.1 & 1941.6 & 2131.1 & 2320.3 & 2507.7 & 2698.9 & 2887.8 \\ 
$z=2.00$ & 435.4 & 621.7 & 809.1 & 996.8 & 1186.1 & 1374.5 & 1563.0 & 1752.1 & 1941.6 & 2131.1 & 2320.3 & 2507.7 & 2698.9 & 2887.8 \\ 
$z=2.10$ & 435.4 & 621.7 & 809.1 & 996.8 & 1186.1 & 1374.5 & 1563.0 & 1752.1 & 1941.6 & 2131.1 & 2320.3 & 2507.7 & 2698.9 & 2887.8 \\ 
$z=2.20$ & 440.0 & 627.5 & 816.5 & 1005.4 & 1195.8 & 1385.3 & 1574.9 & 1765.1 & 1955.8 & 2146.8 & 2337.4 & 2526.2 & 2718.2 & 2908.5 \\ 
$z=2.30$ & 443.4 & 635.7 & 829.7 & 1023.7 & 1218.5 & 1412.5 & 1608.0 & 1803.2 & 1997.8 & 2194.1 & 2390.2 & 2585.2 & 2779.8 & 2976.9 \\ 
$z=2.40$ & 522.9 & 754.6 & 986.8 & 1220.4 & 1453.9 & 1688.1 & 1921.9 & 2155.9 & 2390.2 & 2624.2 & 2857.2 & 3091.9 & 3327.0 & 3558.6 \\ 
$z=2.50$ & 622.6 & 907.0 & 1193.1 & 1479.5 & 1765.1 & 2052.7 & 2340.0 & 2627.1 & 2914.0 & 3200.1 & 3487.7 & 3776.2 & 4062.2 & 4350.0 
\enddata
\label{tab:exposuretime}
\end{deluxetable*}

\end{document}